\documentclass[%
 aip,
 amsmath,amssymb,
 preprint,%
]{revtex4-1}

\draft
\usepackage{graphicx}
\usepackage{braket}
\usepackage{bm}
\usepackage{amsmath}
\newcommand{\kbt}{\beta_\text{T}}
\newcommand{\pcm}{\text{ cm}^{-1}}

\begin{document}

\title{Accuracy of approximate methods for the calculation of absorption-type linear spectra with a complex system-bath coupling} 

\author{J.A. N\"{o}thling}

\affiliation{Department of Physics, University of Pretoria, 0002 Pretoria, South Africa}
\affiliation{National Institute for Theoretical and Computational Sciences (NITheCS), South Africa}
\author{Tom\'{a}\v{s} Man\v{c}al}
\affiliation{Faculty of Mathematics and Physics, Charles University, Ke Karlovu 5, CZ-121 16 Prague 2, Czech Republic}
\author{T.P.J. Kr\"{uger}}
\affiliation{Department of Physics, University of Pretoria, 0002 Pretoria, South Africa}
\affiliation{National Institute for Theoretical and Computational Sciences (NITheCS), South Africa}

\begin{abstract}
The accuracy of approximate methods for calculating linear optical spectra depends on many variables. In this study, we fix most of these parameters to typical values found in photosynthetic light-harvesting complexes of plants and determine the accuracy of approximate spectra with respect to exact calculation as a function of the energy gap and interpigment coupling in a pigment dimer. We use a spectral density with the first eight intramolecular modes of chlorophyll \textit{a} and include inhomogeneous disorder for the calculation of spectra. We compare the accuracy of absorption, linear dichroism, and circular dichroism spectra calculated using the Full Cumulant Expansion (FCE), coherent time-dependent Redfield (ctR), and time-independent Redfield and modified Redfield methods. As a reference we use spectra calculated with the Exact Stochastic Path Integral Evaluation method. We find the FCE method to be the most accurate for the calculation of all spectra. The ctR method performs well for the qualitative calculation of absorption and linear dichroism spectra when pigments are moderately coupled ($\sim 15\pcm$), but ctR spectra may differ significantly from exact spectra when strong interpigment coupling ($\sim 100\pcm$) is present. The dependence of the quality of Redfield and modified Redfield spectra on molecular parameters is similar, and these methods almost always perform worse than ctR, especially when the interpigment coupling is strong or the excitonic energy gap is small (for a given coupling). The accuracy of approximate spectra is not affected by resonance between the excitonic energy gap and intramolecular modes when realistic inhomogeneous disorder is included.
\end{abstract}

\pacs{}

\maketitle 

\section{Introduction}
Linear optical spectroscopy is a crucial analytical tool in biology, chemistry, materials
science, molecular physics, and various other disciplines. It provides information about, i.a., the dipole strengths and orientations\cite{Garab2009, rodger1997circular, Romero2009}, energy landscape\cite{Sznee2004, Kruger2010, Novoderezhkin2004a, Ramanan2015}, exciton structure\cite{Palacios2002, Bulheller2007}, and stoichiometry\cite{Croce2002} in pigment aggregates. In principle, all the information contained in an experimental spectrum could be extracted by comparison with exact simulations. The exact calculation of linear spectra is possible for Gaussian environmental fluctuations\cite{Tanimura1989}, but is computationally expensive, and is generally only performed in order to benchmark approximate calculations on simple systems. One reliable approach for the calculation of exact spectra is the Hierarchical Equations of Motion (HEOM) method\cite{Tanimura1989, Ishizaki2009}, which requires solving many tiers of equations for the system's reduced density matrix and is therefore memory intensive\cite{Kreisbeck2014}. The addition of a single high-frequency mode to the environment's spectral density greatly increases the number of equations that have to be solved\cite{Novoderezhkin2015}. An alternative approach is the method of Exact Stochastic Path Integral Evaluation (PI), which is easier to implement than HEOM, converges faster, and is not memory intensive\cite{Moix2015}. Nevertheless, as we discuss in this article, this method also scales poorly with the number of high-frequency modes.

The intractability of exact methods prevents their use for the calculation of ensemble spectra of large, disordered aggregates with complex system--environment coupling. Many different approaches to the approximate calculation of linear optical spectra exist\cite{Renger2002, Valkunas2013, Ma2015, Gelzinis2015}. Most of these methods rely on the second-order cumulant expansion of the system--environment coupling, and they differ mainly in their treatment of the off-diagonal elements of this coupling in the exciton basis. In two common and economical approaches, the time-independent Redfield and modified Redfield rates are included phenomenologically in the calculation of linear spectra to account for the lifetime broadening of electronic transitions\cite{Novoderezhkin2004a, Gelzinis2015, Jassas2018}. A more explicit treatment of the second-order cumulant expansion brings about the Full Cumulant Expansion (FCE) method\cite{Ma2015, Cupellini2020}, for which the time-dependent expressions are slower to evaluate than the rates of the Redfield-type methods, but are more accurate. Application of the secular approximation to FCE yields the faster, but less accurate, coherent time-dependent Redfield (ctR) method \cite{Gelzinis2015}.

When calculating approximate spectra, it is essential to know what level of accuracy to expect from the method that is used, under which conditions it may be substituted with a more cost-efficient one, and what effect different approximations have on the quality of spectra. Since the approximate methods discussed above are used frequently, rely on different, common, approximations, and vary significantly in speed, their comparison provides valuable insight into the capabilities and limitations of approximate methods in general and how they may be improved. Although some of these approximate methods have been compared with one another, the previous comparisons were either done for a specific system without much variation in parameters\cite{Cupellini2020} or for model systems with unrealistic system--environment coupling\cite{Gelzinis2015, Schroder2006}. Therefore, there is clearly a need for a systematic comparison of approximate methods for calculating linear spectra in a system with realistic system--environment coupling.

Gelzinis \textit{et al.}\cite{Gelzinis2015} compared the accuracy of absorption spectra obtained for a dimer using the ctR, Redfield and modified Redfield methods. They calculated the spectral qualities as functions of various system and environment parameters which they varied one at a time. Inhomogeneous disorder, for instance, was considered as one of their parameters, but was set to zero when other parameters were varied. They found ctR spectra to agree significantly better with exact spectra over a wide range of parameter values than spectra calculated with the Redfield and modified Redfield methods. They also found the modified Redfield method to perform worst for all the parameter values that they considered. For most of their calculations, Gelzinis \textit{et al.} used an overdamped Brownian oscillator spectral density without any underdamped (high-frequency) modes. They included a single high-frequency mode for one of their comparison calculations and treated its energy as the only variable in that comparison. 

Cupellini \textit{et al.}\cite{Cupellini2020} used the FCE and modified Redfield methods to calculate absorption and circular dichroism (CD) spectra of the purple bacterial light-harvesting complex LH2 and compared these spectra to experimental data. The FCE spectra agreed better with experimental spectra than the modified Redfield spectra did. However, because of the uncertainty about the true molecular parameters, none of the spectra agreed well with the experimental spectra, and their comparison is consequently not as conclusive as it would be for a model study. 

In real pigment aggregates, many of the parameters considered by Gelzinis \textit{et al.}, such as the amplitude and shape of the environmental spectral density and the variance of the inhomogeneous disorder, can be estimated accurately and have similar values for a wide range of aggregates. The parameters that typically have the largest variation in the simulation of spectra are the temperature, energies of pigments, and interpigment couplings.

To our knowledge, neither FCE nor ctR has been used for the calculation of linear dichroism (LD) spectra, and the non-secular dipole dependence of LD has not been previously derived. In this article, we compare absorption, LD, and CD spectra calculated with the FCE, ctR, standard Redfield, and modified Redfield methods. We determine the dependence of the spectral accuracy on the simultaneous variation of pigment energy and interpigment coupling at two different temperatures for inhomogeneously broadened absorption-type spectra. Unlike Gelzinis \textit{et al.}\cite{Gelzinis2015}, who used HEOM for their exact calculations, we calculate exact spectra using the more economical PI method\cite{Moix2015}. This allows us to better model environmental interactions by including eight high-frequency modes in our spectral density. In order for our conclusions to be accurate for photosynthetic light-harvesting complexes, we use the parameters of the first eight intramolecular modes of chlorophyll \textit{a}. The inclusion of several high-frequency modes allows us to determine the effect of excitonic energy gap resonance with intramolecular modes on the quality of calculated linear spectra.

\section{Theory}
\label{Introduction: theoretical background}
\subsection{Hamiltonian}
The spectroscopic response from a light-harvesting complex originates from quantum dynamics that are governed by the Hamiltonian
\begin{equation}
    \hat{H} = \hat{H}_{\text{mol}} + \hat{H}_{\text{rad-mat}}(t),
\end{equation}
in which $\hat{H}_{\text{rad-mat}}(t)$ describes the light-matter interaction in the semi-classical approximation (hence the time-dependence of the Hamiltonian) and $\hat{H}_{\text{mol}}$ describes the influence of molecular attributes on the dynamics.

We treat the pigment aggregate as an open quantum system and partition $\hat{H}_{\text{mol}}$ into a system, bath, and system-bath interaction part:
\begin{equation}
    \label{H_mol}
    \hat{H}_{\text{mol}} = \hat{H}_{\text{S}} + \hat{H}_{\text{B}} + \hat{H}_{\text{SB}}.
\end{equation}
For a molecular system composed of multiple interacting molecules with optical band gaps, the electronic ground state is the collective electronic state in which all molecules are found in their respective electronic ground states.  Optical properties of the system are characterized by transitions to the so-called \textit{single exciton} band, which is composed of collective electronic states with an excitation residing on a single molecule of the system. For a light-harvesting complex with $N$ pigment molecules, the ground state $\ket{g}$ can thus be represented by a tensor product of molecular ground states $\ket{g} = \prod\limits_{n=1}^N \ket{g_n}$, where $\ket{g_n}$ is the ground state of pigment $n$. There are $N$ collective singly excited states represented by $\ket{n} = \prod\limits_{\substack{m=1 \\ m\neq n }}^N \ket{g_m}\ket{\epsilon_n}$, where $\ket{\epsilon_n}$ denotes the electronically excited state of pigment $n$.
The system Hamiltonian, the so-called Frenkel exciton Hamiltonian, involves the energies $\varepsilon_n$ of these collective states and their mutual interaction energies $J_{mn}$. It has the general form

\begin{equation}
    \label{Hamiltonian in site basis}
    \hat{H}_S = \sum\limits_{n=1}^N \varepsilon_n\Ket{n}\Bra{n} + \sum\limits_{n=1}^N \sum\limits_{\substack{m=1 \\ m\neq n}}^N J_{mn}\Ket{m}\Bra{n},
\end{equation}
where the ground-state energy is set to zero and $\varepsilon_n$ is the energy of the vertical (Franck-Condon) transition on pigment $n$.

Electronic transitions observed by spectroscopic means occur between the ground state and excited eigenstates of the Hamiltonian in Eq. (\ref{Hamiltonian in site basis}). These eigenstates can be found by performing an affine \textit{exciton} transformation on $\hat{H}_S$, which renders it in the simple form:
\begin{equation}
    \label{Hamiltonian in exciton basis}
    \hat{H}_S = \sum\limits_{\alpha=1}^N \varepsilon_\alpha\Ket{\alpha}\Bra{\alpha},
\end{equation}
where the exciton states are linear combinations of the site excitations, i.e., $\Ket{\alpha} = \sum\limits_{n=1}^N c^{n\alpha}\Ket{n}$, and $\varepsilon_\alpha$ represent the eigenenergies of the Hamiltonian.

The exciton energies are modulated by nuclear fluctuations in the pigment molecules and their environment and are therefore coupled to phonon modes of the environment and intramolecular vibrational modes. These modes together constitute the \textit{bath} and are modeled as an infinite set of independent harmonic oscillators,
\begin{equation}
    \label{H_B}
    \hat{H}_{\text{B}} = \sum\limits_{k=1}^\infty\frac{\omega_k}{2}(\bm{\hat{p}}_k^2 + \bm{\hat{q}}_k^2)\otimes \hat{I}_S,
\end{equation}
where $\hat{I}_S$ is the identity operator in the system's Hilbert space, and $\omega_k$, $\bm{\hat{p}}_k$, and $\bm{\hat{q}}_k$ are the frequency and position and momentum operators of the $k^{\text{th}}$ mode, respectively. In this article, we use units such that $\hbar=1$.
Typically, the system--bath coupling is assumed to be linear:
\begin{equation}
    \label{H_SB}
    \hat{H}_{\text{SB}} = -\sum\limits_{k=1}^\infty\sum\limits_{n=1}^N\omega_k d_{k,nn} \hat{q}_k\Ket{n}\Bra{n},
\end{equation}
and its strength is characterized by the reorganization energy  
\begin{equation}
    \label{lambda}
    \lambda_n = \sum\limits_{k=1}^\infty \frac{\omega_k}{2} d^2_{k,nn}
\end{equation}
associated with each electronic transition. The reorganization energy can often be estimated from experimental measurements (e.g. from the Stokes shift) and depends on the dimensionless shift $d_{k,nn}$ of the excited state harmonic potential energy surface with respect to the ground state potential energy surface.  In Eq. (\ref{H_SB}), spatial dimensions were separated so that $\hat{q}_k$ is a scalar operator.

The light--matter interaction is treated through linear coupling to a classical electric field $\bm{E}(t)$:
\begin{equation}
    \hat{H}_{\text{rad-mat}}(t) = -\bm{\hat{\mu}}\cdot\bm{E}(t),
\end{equation}
with the transition dipole moment operator given by
\begin{equation}
\label{transition dipole moment}
   \bm{\hat{\mu}} = \sum\limits_{n=1}^N \bm{{\mu}}_n(\Ket{n}\Bra{g} + \Ket{g}\Bra{n}),
\end{equation}
where the vector $\bm{{\mu}}_n$ is the transition dipole moment of the transition from the ground to the excited state on the $n^\text{th}$ molecule. The dipole moments are assumed to be independent of the nuclear coordinates (i.e., the Condon approximation is made).

\subsection{Linear spectra}
During linear spectroscopic measurements, the incident electromagnetic field induces a transverse molecular polarization, which generates an additional electric field interfering with the incident fields. The combination of these fields is then measured as a field leaving the macroscopic molecular sample. The linear response (first-order perturbation of the system) is characterized by the dipole-dipole correlation function 
\begin{equation}
    \label{dipole-dipole correlation}
    \langle \mu(t)\mu(0)\rangle = \text{Tr}[e^{i\hat{H}_{\text{mol}}t}  \bm{\hat{\mu}} e^{-i\hat{H}_{\text{mol}}t} \bm{\hat{\mu}} \hat{\rho}_{\text{eq}}],
\end{equation}
where $\text{Tr}[\,\bullet\,]$ denotes the trace operation and $\hat{\rho}_{\text{eq}}$ is the equilibrium density matrix.

We assume that the pigment aggregate is in thermal equilibrium before it is excited with light and that the equilibrium state is separable, with the system in its collective electronic ground state, i.e., $\hat{\rho}_{\text{eq}} = \ket{g}\bra{g}\hat{\rho}^{\text{eq}}_B$. With this equilibrium state, it is straightforward to show that the dipole-dipole correlation function in Eq. (\ref{dipole-dipole correlation}) can be written as
\begin{eqnarray}
    \langle \mu(t)\mu(0)\rangle = \sum\limits_{m=1}^N \sum\limits_{n=1}^N  &&\bm{\mu}_m \cdot \bm{\mu}_n \text{Tr}_B [\bra{g}e^{i\hat{H}_{\text{mol}}t}\ket{g} \nonumber \\  &&\times \bra{m}e^{-i\hat{H}_{\text{mol}}t}\ket{n} \hat{\rho}_B^{\text{eq}}] \label{dipole-dipole correlation expanded}.
\end{eqnarray}
Based on Eq. (\ref{dipole-dipole correlation expanded}), we define the absorption tensor as
\begin{equation}
    \label{Absorption tensor}
    \text{I}^{\text{A}}(t) = \text{Tr}_B[e^{i\hat{H}_g t}e^{-i\hat{H}_{e} t}\hat{\rho}^{\text{eq}}_B],
\end{equation}
where $\hat{H}_g$ and $\hat{H}_{e}$ are the ground-state and excited-state block of $\hat{H}_{\text{mol}}$, respectively.
Absorption spectra $\text{S}^{\text{A}}(\omega)$ are obtained by the Fourier transform of the dipole-dipole correlation function (Eq. \ref{dipole-dipole correlation}). In terms of the absorption tensor (Eq. (\ref{Absorption tensor})), we obtain
\begin{equation}
    \label{Absorption spectrum}
    \text{S}^{\text{A}}(\omega) \propto \omega\sum\limits_{m=1}^N \sum\limits_{n=1}^N f^{\mu, \text{A}}_{mn}\,\Bigl[2\text{Re}\int\limits_{0}^{\infty}dt\, e^{i\omega t} \text{I}^{\text{A}}_{mn}(t)\Bigr], 
\end{equation}
with the dipole factor $f^{\mu, \text{A}}_{mn} = (\bm{\mu}_m\cdot\bm{\mu}_n)$.

In a similar way as for the absorption spectrum, the CD spectrum can also be formulated based on a correlation function, this time being the electric dipole--magnetic dipole correlation function.  The spectrum $\text{S}^{\text{CD}}(\omega)$ is given by the right-hand side of Eq. (\ref{Absorption spectrum}) with the dipole factor\cite{Cupellini2020}:
\begin{equation}
    \label{CD dipole factor}
    f^{\mu, \text{CD}}_{mn} = \sqrt{\varepsilon_m\varepsilon_n}(\bm{R}_m - \bm{R}_n)\cdot(\bm{\mu}_m \times \bm{\mu}_n).
\end{equation}

The LD spectrum, $\text{S}^{\text{LD}}(\omega)$, is also obtained from the right-hand side of Eq. (\ref{Absorption spectrum}). As shown in Appendix \ref{Dipole factor for linear dichroism of disk-shaped complexes}, the dipole factor for LD measurements on disk-shaped pigment aggregates is given by
\begin{equation}
    \label{LD dipole factor}
    f^{\mu, \text{LD}}_{mn} = \bm{\mu}_m\cdot\bm{\mu}_n - 3|\bm{\mu}_m||\bm{\mu}_n|\cos\alpha_m\cos\alpha_n,
\end{equation}
where $\alpha_m$ is the angle between $\bm{\mu_m}$ and the (uniquely chosen) vector normal to the disk.

\subsection{Absorption tensor}
By using an operator identity, Eq. (\ref{Absorption tensor}) can be written in the time-ordered exponential form as\cite{Mukamel1995}
\begin{equation}
    \label{Absorption tensor with time-ordered exp i.t.o. H_eg}
    \text{I}^{\text{A}}(t) = \text{Tr}_B\Bigl[\exp_+\bigl[{-i\int_0^t d\tau\, \hat{H}_{eg}(\tau)}\bigr]\hat{\rho}^{\text{eq}}_B\Bigr],
\end{equation}
with $\hat{H}_{eg}(\tau) = e^{i\hat{H}_g\tau}(\hat{H}_{e} - \hat{H}_g)e^{-i\hat{H}_g\tau}$.
Using Eqs. (\ref{H_mol}), (\ref{H_B}), and (\ref{H_SB}), we can rewrite Eq. (\ref{Absorption tensor with time-ordered exp i.t.o. H_eg}) as
\begin{equation}
    \label{Absorption tensor with time-ordered exp i.t.o. H_SB}
    \text{I}^{\text{A}}(t) = \text{Tr}_B\Bigl[\exp_+\bigl[{-i\bigl(\hat{H}_{S}t + \int_0^t d\tau\, \hat{H}_{SB}(\tau)}\bigr)\bigr]\hat{\rho}^{\text{eq}}_B\Bigr],
\end{equation}
where $\hat{H}_{SB}(\tau) = e^{i\hat{H}_g\tau}\hat{H}_{SB}e^{-i\hat{H}_g\tau}$.
When the system-bath coupling is linear and the bath is harmonic, all the information about the bath that is necessary for the calculation of spectra (both linear and nonlinear) is contained in the correlation function of $\hat{H}_{SB}(t)$:
\begin{equation}
    \label{correlation function}
    C_{n}(t) = \sum\limits_{k=1}^\infty \omega_k^2 d^2_{k, nn}\Bigl\langle \hat{q}_k(t) \hat{q}_k(0) \Bigr\rangle.
\end{equation}
The correlation function can be determined from its corresponding spectral density $C_n(\omega)$ (see e.g. Ref. \onlinecite{Mukamel1995}) as
\begin{equation}
    C_n(t) = \frac{1}{\pi} \int_0^\infty d\omega \,C_n(\omega) \frac{\cosh(\kbt\omega/2 - i\omega t)}{\sinh(\kbt \omega/2)},
\end{equation}
where $\kbt = \frac{1}{k_B T}$ with $k_B$ the Boltzmann constant and $T$ the temperature. The spectral density may be modelled straightforwardly with experimentally motivated parameters (see Section \ref{Calculations}).

\subsubsection{Full cumulant expansion (FCE)}
\label{Full cumulant expansion (FCE)}
Since the trace, $\text{Tr}_B\bigl[\bullet \, \hat{\rho}^{\text{eq}}\bigr]$, amounts to the calculation of a statistical average, it is sensible to expand Eq. (\ref{Absorption tensor with time-ordered exp i.t.o. H_SB}) in the cumulants of $\hat{H}_{SB}(t)$. Up to second order, such a cumulant expansion yields\cite{Ma2015}
\begin{equation}
    \label{FCE absorption tensor}
    \text{I}^{\text{A}}(t) = e^{-iH_St}e^{-K(t)}
\end{equation}
with
\begin{eqnarray}
    K_{\alpha\beta}(t) &= \sum\limits_{\delta=1}^N \int_0^t d\tau
     \int_0^\tau d\tau'\, e^{i\omega_{\alpha\delta}\tau- i\omega_{\beta\delta}\tau'} \nonumber \\ 
     &\times  C_{\alpha\delta\delta\beta}(\tau - \tau')\label{FCE K}.
\end{eqnarray}
If the energy gap fluctuations on different molecules are independent and the correlation functions of all the pigments have the same form (i.e., $C_n(t)=C(t)$ $\forall \, n$), we may write
\begin{equation}
    \label{C participation corrected}
    C_{\alpha\delta\delta\beta}(t) = \gamma_{\alpha\delta\delta\beta}C(t),
\end{equation}
with $\gamma_{\alpha\delta\delta\beta} = \sum\limits_{n=1}^Nc^{n\alpha}c^{n\delta}c^{n\delta}c^{n\beta}$.

It is important to note that the cumulant expansion is a nested expansion; even though only the first two cumulants (mean and variance) are included in the expressions above, the Taylor expansion of $\exp(\,\bullet\,)$ involves all orders of $H_{SB}$. When a distribution is fully described by its first two cumulants (as it is the case for a Gaussian distribution), the cumulant expansion may be (but need not be) exact.

\subsubsection{Coherent time-dependent Redfield theory (ctR)}
\label{Coherent time-dependent Redfield theory}
The calculation of Eq. (\ref{FCE absorption tensor}) can be greatly simplified by making the \textit{secular approximation} and requiring the matrix elements $\Bra{\alpha}e^{-i\hat{H}_{e} t}\Ket{\beta}$ to be zero for $\alpha\neq\beta$.
The cumulant expansion then yields for the absorption tensor\cite{Gelzinis2015}
\begin{equation}
     \label{coherent time-dependent Redfield absorption}
     \text{I}_{\alpha\beta}^{\text{A}}(t) = \delta_{\alpha\beta}e^{-i\varepsilon_\alpha t - g_{\alpha\alpha\alpha\alpha}(t) - \xi_\alpha(t)},
\end{equation}
with
\begin{equation}
    \label{lineshape function}
    g_{\alpha\alpha\alpha\alpha}(t) = \int_0^t d\tau \int_0^\tau d\tau'\, C_{\alpha\alpha\alpha\alpha}(\tau')
\end{equation}
and 
\begin{equation}
    \label{ctR rate term}
    \xi_\alpha(t) = \sum\limits_{\beta \neq \alpha}^N \int_0^t d\tau \int_0^\tau d\tau'\, e^{i\omega_{\alpha\beta } \tau'}C_{\alpha\beta\beta\alpha}(\tau'). 
\end{equation}
In Eq. (\ref{coherent time-dependent Redfield absorption}), $g_{\alpha\alpha\alpha\alpha}(t)$ describes pure dephasing, and the real part of $\xi_{\alpha}(t)$ describes lifetime broadening (see Fig. S1 and the discussion in Sections \ref{Dependence of spectral quality on site energy and excitonic coupling} and \ref{Dependence of spectral quality on the excitonic energy gap}). The imaginary part of $\xi_{\alpha}(t)$ describes modulation of the exciton energies due to the coupling between exciton states by the off-diagonal system-bath coupling elements.

\subsubsection{Standard and modified Redfield theories}
\label{Standard and modified Redfield theories}
As stated in the previous section, the term $\xi_\alpha(t)$ in Eq. (\ref{coherent time-dependent Redfield absorption}) accounts for dephasing due to energy relaxation. The same effect can be achieved, phenomenologically, by substituting  $\xi_\alpha(t)$ with $R_\alpha t$ for $R_\alpha$ a time-independent rate constant calculated as
\begin{equation}
    R_{\alpha} = \frac{1}{2}\sum\limits_\beta R_{\beta\alpha},
\end{equation}
where $R_{\beta\alpha}$ is calculated using standard time-independent Redfield theory\cite{Valkunas2013}:
\begin{equation}
    \label{Redfield rate}
    R^{\text{sR}}_{\beta\alpha} = 2\text{Re}\,\int_{0}^\infty dt\, e^{i\omega_{\alpha\beta}t}C_{\alpha\beta\beta\alpha}(t),
\end{equation}
or time-independent modified Redfield theory\cite{Zhang1998}:
\begin{eqnarray}
R^{\text{mR}}_{\beta\alpha} = && 2\text{Re}\,\int_{0}^\infty dt \,e^{i\omega_{\alpha\beta}t} \nonumber\\
&& \times \exp \Bigl(-g_{\beta\beta\beta\beta}(t) - g_{\alpha\alpha\alpha\alpha}(t) + g_{\alpha\alpha\beta\beta}(t) \nonumber\\
&& + g_{\beta\beta\alpha\alpha}(t) - 2i(\lambda_{\alpha\alpha\alpha\alpha} - \lambda_{\beta\beta\alpha\alpha})t\Bigr)  \nonumber\\
&& \times \Bigl( C_{\beta\alpha\alpha\beta}(t) - [\dot{g}_{\alpha\beta\beta\beta}(t) - \dot{g}_{\alpha\beta\alpha\alpha}(t) - 2i\lambda_{\alpha\beta\alpha\alpha}]  \nonumber\\
&& \times [\dot{g}_{\beta\beta\beta\alpha}(t) - \dot{g}_{\alpha\alpha\beta\alpha}(t) - 2i\lambda_{\beta\alpha\alpha\alpha}]\Bigr)\label{modified Redfield rate}.
\end{eqnarray}
with $\lambda_{\alpha\beta\gamma\delta}=\sum_{n=1}^{N}c^{n\alpha}c^{n\beta}c^{n\gamma}c^{n\delta}\lambda$ analogous to Eq. (\ref{C participation corrected}). The reorganization energy can be calculated from the spectral density as $\lambda=\int_{-\infty}^{\infty}\frac{d\omega}{2\pi\omega}C(\omega)$.

Notice that the upper limit of integration in Eqs. (\ref{Redfield rate}) and (\ref{modified Redfield rate}) is $t\rightarrow\infty$, and these expressions therefore describe the Markovian long-time Redfield and modified Redfield rates. 

As in Section \ref{Coherent time-dependent Redfield theory}, the secular approximation was invoked in the derivation of the Redfield and modified Redfield rates above. It is well known that the non-secular Redfield relaxation tensor may lead to positivity breakdown in the excited state population dynamics, while the secular form in Eq. (\ref{Redfield rate}) (as the so-called Lindbladian) yields strictly positive populations. When calculating linear spectra, the non-positivity may not be problematic, since the coherence block of the density matrix, which takes part in the evaluation of the linear spectra, evolves independently of the block of excited state dynamics. Although often quantitatively incorrect, the non-secular Redfield rates shows the right qualitative behaviour in describing non-secular effects in the linear spectra \cite{Olsina2014}. The secular approximation may significantly decrease the size of the computational problem, especially for large pigment aggregates, however, and is therefore used in this study.

\subsubsection{Exact stochastic path integral evaluation (PI)}
\label{Exact stochastic path integral evaluation}
It may well be impossible to find an exact closed expression for the absorption tensor. Exact calculations can, however, be performed by Monte Carlo path integration\cite{Moix2015}. In this method, Eq. (\ref{Absorption tensor with time-ordered exp i.t.o. H_SB}) is considered as the average (indicated by $\text{Tr}_B[\,\bullet\,]$) of many solutions of the equation
\begin{equation}
    \label{Differential equation for stochastic calculation of Absorption}
    \frac{d}{dt}\rho^{\text{A}}(t) = -i\Bigl(\hat{H}_S + \sum\limits_{n=1}^\infty \xi_n(t)\Ket{n}\Bra{n}\Bigr)\rho^{\text{A}}(t),
\end{equation}
subject to the initial condition $\rho^{\text{A}}(0) = \hat{I}$, where $\hat{I}$ is the identity operator in the system's Hilbert space. 
In Eq. (\ref{Differential equation for stochastic calculation of Absorption}), $\xi_n(t)$ is a stochastic process conforming to the statistics of $\hat{H}_{SB}(t)$, namely,
\begin{eqnarray}
    \langle\xi_m(t)\xi_n(t')\rangle &&= \delta_{mn} C_n(t-t') \nonumber\\
    \langle\xi_n(t)\rangle &&= 0.
\end{eqnarray}
The absorption tensor is calculated as the average of many stochastic solutions of Eq. (\ref{Differential equation for stochastic calculation of Absorption}):
\begin{equation}
    \text{I}^{\text{A}}(t) = \langle \rho^{\text{A}}(t) \rangle_\xi.
\end{equation}
As shown in Ref. \onlinecite{Moix2015}, this average converges to the exact absorption tensor as the number of stochastic trajectories increases to infinity.

\section{Calculations}
\label{Calculations}

For the calculations in this study we consider a molecular dimer with a single electronic transition on each molecule. For the sake of computational convenience we subtract the energy of the lower frequency electronic transition from the energies of the excited states, which amounts to applying the so-called Rotating Wave Approximation (RWA). This allows us to reduce the problem to just the treatment of the excited state block of the Hamiltonian, which then takes the form
\begin{equation}
    \label{Hamiltonian for calculations}
    H_S = \begin{bmatrix}
       \epsilon & J\\
       J & 0
    \end{bmatrix},
\end{equation}
where $\epsilon$ is the site basis energy gap between the electronic transitions on the two molecules of the dimer, and $J$ is their resonance coupling.  

The exact method of stochastic path integration  (PI; see Section \ref{Exact stochastic path integral evaluation}) is significantly more tractable for the calculation of linear spectra than using the Hierarchical Equations of Motion (HEOM), as was done in earlier works\cite{Gelzinis2015, Novoderezhkin2015}. Nevertheless, the convergence of PI calculations slows precipitously as more high-frequency modes are added to the bath. In order for PI convergence to be sufficient, as well as for our conclusions to be valid for real systems, we describe the bath by a spectral density with contributions from a continuum of phonon modes and eight intramolecular vibrations. We assume that all the pigment molecules have the same correlation function. The low-frequency phonon contribution is modeled as a quantum Brownian oscillator with parameters used in earlier studies\cite{Novoderezhkin2004a, Mascoli2020}:
\begin{equation}
    C_{\text{ph}}(\omega) = 2\lambda_{\text{ph}}\frac{\omega\gamma_{\text{ph}}}{\omega^2 + \gamma_{\text{ph}}^2},
\end{equation}
with the reorganization energy $\lambda_{\text{ph}} = 40 \text{ cm}^{-1}$ and spectral width $\gamma_{\text{ph}} = 40 \text{ cm}^{-1}$.
The high-frequency part of the spectral density is implemented as the sum of underdamped modes\cite{Jassas2018} with parameters of the first eight high-frequency modes of the plant light-harvesting complex LHCII\cite{Peterman1997} (see Table \ref{tab:parameters of LHCII spectral density}), which are predominantly due to chlorophyll \textit{a} vibrations of the lowest-energy exciton:

\begin{equation}
    \label{underdamped spectral density}
    C_{\text{hf}}(\omega) = \sum\limits_{i=1}^8 \frac{\omega_i|\omega_i|S_i\gamma_i}{2\bigl((|\omega| - \omega_i)^2 + (\frac{\gamma_i}{2})^2\bigr)},
\end{equation}
with $\gamma_i = 6 \pcm \, \forall \, i$, and the central frequencies ($\omega_i$) and Huang-Rhys factors ($S_i$) given in Table \ref{tab:parameters of LHCII spectral density}.
\begin{table*}
\caption{\label{tab:parameters of LHCII spectral density}Central frequency $\omega$ and Huang-Rhys factor $S$ for each of the underdamped modes used in this study. These modes correspond to the the first eight high-frequency modes of LHCII as determined from fluorescence line-narrowing spectra\cite{Peterman1997}. The Huang-Rhys factors were scaled by a factor of 0.8 to better fit experimental spectra\cite{Novoderezhkin2004a}.}
\begin{ruledtabular}
\begin{tabular}{lllllllll}
$\bm{\omega_i \: (\text{cm}^{-1})}$ & 97 & 138 & 213 & 260 & 298 & 342 & 388 & 425 \\
\hline
$\bm{S_i}$&0.0192 & 0.0230 &0.0240& 0.0214& 0.0214& 0.0483 &0.0199 &0.0119\\
\end{tabular}
\end{ruledtabular}
\end{table*}

Approximate linear spectra (i.e., spectra calculated using the FCE, ctR, Redfield or modified Redfield methods) were calculated as described in Section \ref{Introduction: theoretical background}. To incorporate disorder, site energies were drawn independently from a normal distribution with a FWHM of $140 \pcm$ (a value that is often used in literature for light-harvesting complexes\cite{Jurinovich2015, Muh2014, Jennings2003}) and centered around the applicable energies. An ensemble size of $5000$ was used for disordered calculations of approximate absorption-type spectra. 

For the PI method, disorder sampling commutes with Monte Carlo integration\cite{Moix2015}. The noise trajectories were calculated using the procedure described by Moix \textit{et al.}\cite{Moix2015}, and the stochastic integration was performed with the second-order weak technique described in Appendix \ref{Second order weak scheme}. We found the second-order weak technique significantly more accurate than the first-order strong Milstein method (weak and strong approximations guarantee convergence of the statistical moments and of individual trajectories, respectively, to a certain order---see Kloeden and Platen\cite{kloeden2011numerical} for an extensive discussion of stochastic differential equations and their solution).

We evaluated approximate spectra by comparing them to the exact spectra determined from PI, using the quality factor defined by Gelzinis \textit{et al.} \cite{Gelzinis2015} as a measure of the correctness of an approximate spectrum:
\begin{equation}
    \label{quality factor}
    Q =  \frac{\int_{-\infty}^\infty d\omega\bigl(S^{\text{PI}}(\omega)\cap S^{\text{approximate}}(\omega)\bigr)}{\int_{-\infty}^\infty d\omega\bigl(S^{\text{PI}}(\omega)\cup S^{\text{approximate}}(\omega)\bigr)},
\end{equation}
where $S^{\text{PI}}$ and $S^{\text{approximate}}$ are the normalized spectra obtained from PI and an approximate method, respectively. In Eq. (\ref{quality factor}), $\cap$ and $\cup$ denote the intersection and union of spectra, respectively. 

All the absorption-type spectra (OD, LD, and CD) can be obtained from the absorption tensor $\text{I}^{\text{A}}(t)$ by element-wise multiplication with the relevant symmetrical dipole factor matrix $f_{mn}$  (see Eqs. (\ref{Absorption spectrum}), (\ref{CD dipole factor}), and (\ref{LD dipole factor})). Since the spectra were normalized when calculating quality, only the ratio of the dipole matrix elements is important. For simplicity, we omitted the frequency and energy prefactors in Eqs. (\ref{Absorption spectrum}) and (\ref{CD dipole factor}) when calculating spectra. For the most part, we consider the following four dipole factor matrices in our calculations: 
\begin{eqnarray}
   &f^{1,0}_{0,1} = \begin{bmatrix}
   1 & 0\\
   0 & 1
\end{bmatrix}, \hspace{5mm}
&f^{1,1}_{1,1} = \begin{bmatrix}
   1 & 1\\
   1 & 1
\end{bmatrix}, \\ \nonumber
&f^{1,-1}_{-1,0} = \begin{bmatrix}
   1 & -1\\
   -1 & 0
\end{bmatrix}, \hspace{2mm} \text{and} \hspace{2mm}
&f^{0,1}_{1,0}  = \begin{bmatrix}
   0 & 1\\
   1 & 0
\end{bmatrix},
\end{eqnarray}
where $f^{1,0}_{0,1}$ and $f^{1,1}_{1,1}$ correspond to absorption spectra when the two dipole moments of the dimer have equal strength and are orthogonal and parallel to each other, respectively, $f^{1,-1}_{-1,0}$ corresponds to an LD spectrum with $\bm{\mu}_1$ a unit vector orthogonal to the LD axis (see Appendix \ref{Dipole factor for linear dichroism of disk-shaped complexes}), $\bm{\mu}_2$ at the magic angle ($54.7^\circ$) relative to the LD axis, and $\bm{\mu}_1\cdot\bm{\mu}_2=-1$, and $f^{0,1}_{1,0}$ is the dipole factor for CD. 

Note that the secular approximation is almost fully valid for the dipole factor $f^{1,0}_{0,1}$, for which Eq. (\ref{Absorption spectrum}) amounts to a trace over the (Fourier-transformed) absorption tensor. The trace is basis-independent and therefore depends only on the diagonal elements of the secular absorption tensor in the exciton basis. Transfer between these diagonal tensor elements is calculated secularly for the ctR, Redfield, and modified Redfield methods but corresponds very well with interdiagonal transfer for the nonsecular FCE method, as shown in Section \ref{Dependence of spectral quality on the excitonic energy gap}.

\section{Results and discussion}
\label{Results and discussion}
\subsection{Dependence of spectral quality on site energy gap and excitonic coupling}
\label{Dependence of spectral quality on site energy and excitonic coupling}
\begin{figure*}[!htbp]
\includegraphics[scale=1.0]{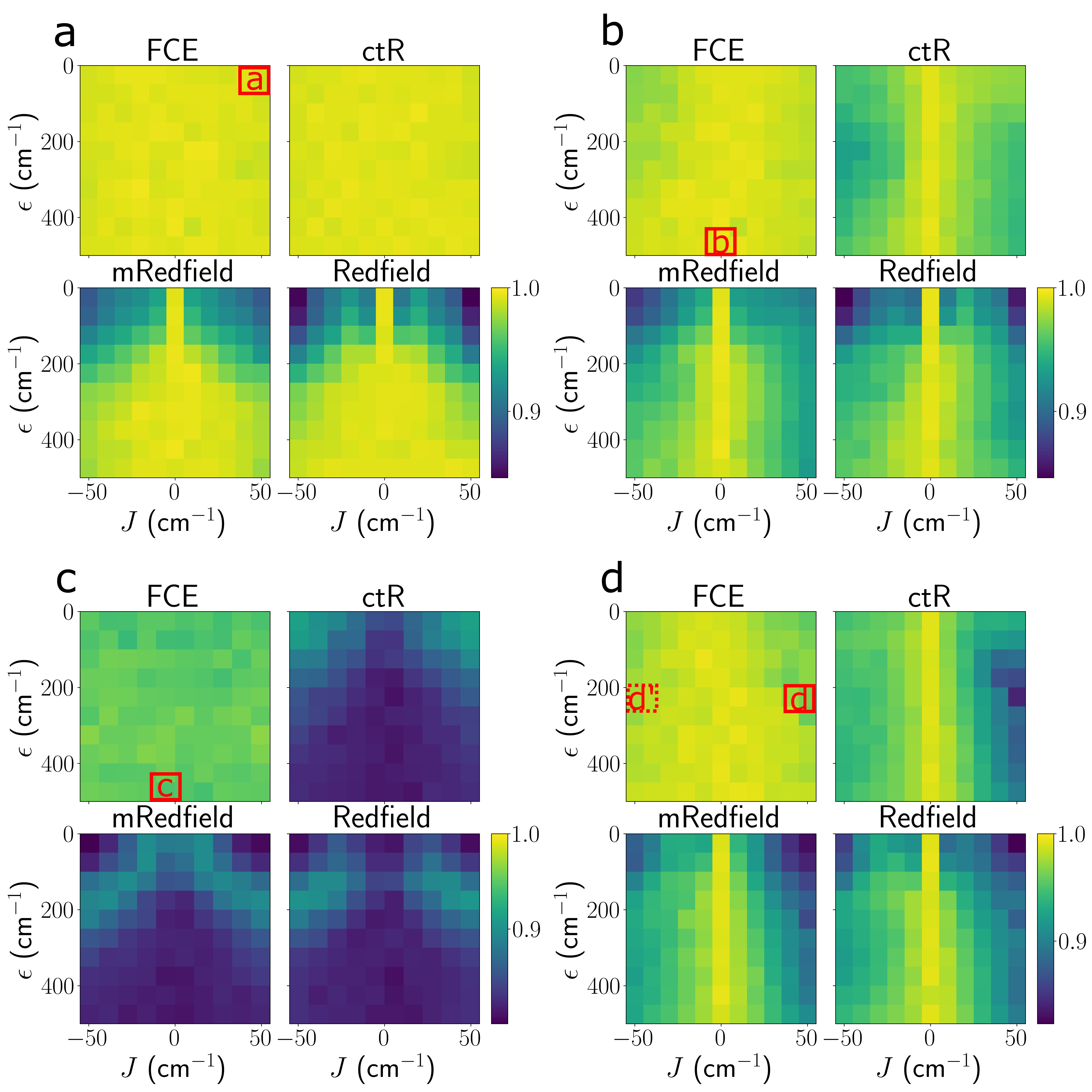}
\caption{\label{fig:energy coupling grid absorption 300K} Dependence of the quality of absorption-type spectra on the site energy gap $\epsilon$ and coupling $J$ at $300 \text{ K}$. Qualities are shown for the absorption dipole factors (\textbf{a}) $f^{1,0}_{0,1}$ and (\textbf{b}) $f^{1,1}_{1,1}$, (\textbf{c}) the CD dipole factor $f^{0,1}_{1,0}$, and (\textbf{d}) the LD dipole factor $f^{1,-1}_{-1,0}$. A disorder of $\sigma_\text{FWHM}=140\pcm$ was used for approximate spectra. The coupling--energy gap pairs for which spectra are shown in Fig. \ref{fig:absorption spectra 300K} are indicated with red frames and the inscribed labels correspond with the respective labels in Fig. \ref{fig:absorption spectra 300K}.}
\end{figure*}

\begin{figure}[!htbp]
\includegraphics[scale=1, trim=0cm 1cm 0 2cm, clip]{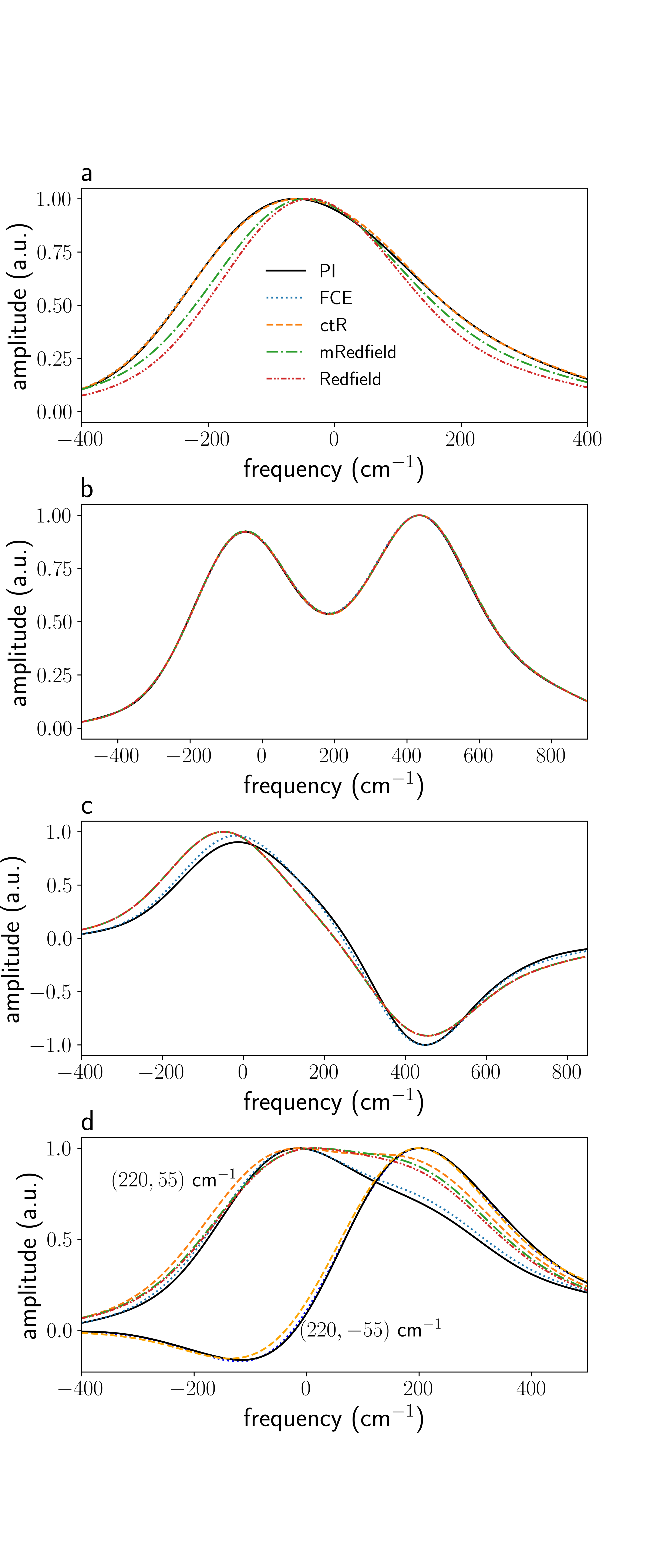}
\caption{\label{fig:absorption spectra 300K} Absorption-type spectra for selected coupling--energy gap pairs and dipole factors, as indicated in Fig. \ref{fig:energy coupling grid absorption 300K}. The same parameters were used for the generation of spectra as were used in Fig. \ref{fig:energy coupling grid absorption 300K}. The spectra for the coupling--energy gap pairs $(55,220)\pcm$ and $(-55,220)\pcm$ refer to points $d$ and $d'$, respectively, in Fig. \ref{fig:energy coupling grid absorption 300K}}
\end{figure}

In order to determine the accuracy of the approximate methods for different Hamiltonians, we varied $\epsilon$ and $J$ in Eq. (\ref{Hamiltonian for calculations}) independently in the ranges $[0,500] \pcm$ and $[-55,55]\pcm$, respectively. The obtained qualities for inhomogeneously broadened absorption-type spectra are shown in Fig. \ref{fig:energy coupling grid absorption 300K} for different dipole factors at $300 \text{ K}$. Spectra of selected coupling--energy gap pairs (indicated for the FCE qualities in Fig. \ref{fig:energy coupling grid absorption 300K}) are shown in Fig. \ref{fig:absorption spectra 300K}. 

It is clear from Fig. \ref{fig:energy coupling grid absorption 300K} that absorption-type spectra calculated with FCE are more accurate than spectra calculated with any of the other methods for the dipole factors in this figure. The quality of these FCE spectra is practically independent of the site energy gap and coupling, and spectra for systems with many pigments (i.e., many different site energies and couplings) may therefore be calculated with predictable quality.

In Section \ref{Full cumulant expansion (FCE)} we noted that the second-order cumulant expansion may be exact when $\hat{q}(t)$ in Eq. (\ref{correlation function}) is a Gaussian process. This expansion indeed leads to exact expressions in the case of a monomeric system (or polymeric system with no interpigment coupling), when the bath is harmonic and $\hat{H}_\text{SB}$ is linear in system and bath operators. Since all of the approximate methods were derived using the second-order cumulant expansion, this perfect correspondence can be seen in Figs. \ref{fig:energy coupling grid absorption 300K}\textbf{a}, \textbf{b} and \textbf{d} for all the methods at zero coupling. The perfect agreement of spectra is also seen in Fig. \ref{fig:absorption spectra 300K}\textbf{b}, where spectra for the different methods are shown for the coupling--energy gap pair $(0, 500)\pcm$ and the dipole factor $f^{1,1}_{1,1}$.  The case for the coupling being exactly zero is not shown for CD qualities (Fig. \ref{fig:energy coupling grid absorption 300K}\textbf{c}) since CD spectra are then identically zero.

For an excitonically coupled system, the cumulant expansion is exact for the treatment of fluctuations induced by the diagonal elements of the system-bath coupling Hamiltonian in the exciton basis. It is not exact for fluctuations induced by the off-diagonal elements, however, since terms that depend on these elements also include excitonic propagators $e^{i\omega_{\alpha\beta}t}$ (see Eqs. (\ref{lineshape function}) and (\ref{ctR rate term}))\cite{Gelzinis2015}. In spite of the second-order truncation, the FCE method is remarkably accurate for the calculation of most spectra in the considered site energy gap and coupling ranges, as seen in Figs. \ref{fig:energy coupling grid absorption 300K}\textbf{a}, \textbf{b} and \textbf{d}. Only for the CD spectrum (Fig. \ref{fig:energy coupling grid absorption 300K}\textbf{c}), when the anti-diagonal dipole factor prescribes significant inclusion of off-diagonal system-bath coupling effects (see Section \ref{Dependence of quality on dipole factor}), and the second-order truncation is therefore least accurate, is the quality of FCE absorption-type spectra notably worse than for the exact approach.

The secular approximation is made in the derivation of the ctR, Redfield and modified Redfield methods but not in the derivation of the FCE method. In fact, this approximation is the only difference between the FCE and ctR methods. For secular methods, the off-diagonal elements of the absorption tensor (Eq. (\ref{Absorption tensor})) are zero in the exciton basis. The discrepancy in quality between FCE and ctR spectra therefore depends on the contribution of these off-diagonal (i.e., nonsecular) elements to the FCE spectra. The spectral contribution from the nonsecular elements depends on both the dipole factor and the system-bath coupling Hamiltonian.

As discussed in Section \ref{Calculations}, the dipole factor $f^{1,0}_{0,1}$ corresponds to the calculation of a nonsecular spectrum, also for secular methods. The isolated effect of the dipole factor on the quality of absorption spectra is seen in Figs. \ref{fig:energy coupling grid absorption 300K}\textbf{a} and \textbf{b}, where the secular approximation is fully valid for Fig. \ref{fig:energy coupling grid absorption 300K}\textbf{a} (see Section \ref{Calculations}), while, for the same dipole strengths, it is least accurate for Fig. \ref{fig:energy coupling grid absorption 300K}\textbf{b}. The dependence of the spectra on the dipole factor also causes the poor quality of the secular CD spectra seen in Fig. \ref{fig:energy coupling grid absorption 300K}\textbf{c}, since the CD dipole factor, $f^{0,1}_{1,0}$, corresponds poorly with $f^{1,0}_{0,1}$. A detailed discussion of the dependence of spectral accuracy on the dipole factor is given in Section \ref{Dependence of quality on dipole factor}.

When the system-bath coupling is treated within the secular approximation, it induces population relaxation but not transfer between exciton populations and coherences, and the off-diagonal elements of the secular absorption tensors therefore remain zero. Such nonsecular transfer dynamics, when considered, would influence the exciton structure by changing the contribution of sites to exciton states in a process called the dynamic localization of excitons\cite{Fassioli2014a}. For the FCE method, nonsecular contributions are due to bath correlation functions of the form $C_{\alpha\beta\beta\beta}$ or $C_{\alpha\alpha\alpha\beta}$. The factor $\gamma_{\alpha\beta\beta\beta}$ defining these correlation functions (see Eq. (\ref{C participation corrected})) is small for small site energy gaps and weak couplings (Fig. S1), and for these parameters the qualities of ctR spectra are therefore almost as high as the qualities of FCE spectra (see Figs. \ref{fig:energy coupling grid absorption 300K}\textbf{b} and \textbf{d}). The quality of CD spectra calculated with the ctR method also corresponds best with FCE qualities at small site energy gaps but (as discussed in Section \ref{Dependence of spectral quality on the excitonic energy gap}) CD spectra depend strongly on nonsecular elements when the coupling is weak, so that the ctR qualities differ significantly from the FCE qualities for weak coupling. The high accuracy of the secular approximation for small site energy gaps and weak couplings is contrary to the expectation\cite{Muh2014, Jassas2018} that the negative effects of neglected dynamic localization on spectral quality are strongest under these conditions. CD spectra calculated with the ctR, Redfield or modified Redfield methods are impacted considerably by both the second-order truncation in the cumulant expansion and by the secular approximation, and their quality is poor for all site energy gaps and couplings considered in Fig. \ref{fig:energy coupling grid absorption 300K}.

The quality of absorption-type spectra calculated using the Redfield method depends similarly on the site energy gap and coupling as the quality of modified Redfield spectra. As shown in  Fig. \ref{fig:energy coupling grid absorption 300K}, these methods yield spectra that are less accurate than FCE and ctR spectra. The difference between the former pair of spectra and ctR spectra (which serve as proxy for the more accurate spectra) is most pronounced at small site energy gaps and strong coupling. The reason for this difference is the energy shift and narrowing of Redfield and modified Redfield spectra relative to ctR spectra, as seen in Fig. \ref{fig:absorption spectra 300K}\textbf{a} for the coupling--energy gap pair $(55,0)\pcm$ and dipole factor $f^{1,0}_{0,1}$. These spectral differences are due to the fact that the Redfield and modified Redfield rates in Eqs. (\ref{Redfield rate}) and (\ref{modified Redfield rate}) are real quantities, whereas the corresponding term, $\xi(t)$, in Eq. (\ref{ctR rate term}) is complex\cite{Gelzinis2015}. Fig. S2 shows the Redfield and long-time ctR rates that were not corrected for pigment participation in excitons (i.e., $\gamma_{\alpha\beta\beta\alpha}$ was set equal to 1 in Eq. (\ref{C participation corrected})). As seen in this figure, the imaginary rate-induced energy splitting in ctR spectra increases with the difference between exciton energies (exciton gap) for small exciton gaps, while the rate-induced spectral shift stays fairly constant at a value of about $30 \pcm$ (see Figs. S2 and \ref{fig:absorption spectra 300K}\textbf{a}). Although the rate-induced splitting in ctR spectra decreases momentarily for exciton gaps larger than about $35\pcm$, the disorder-averaged participation factor $\gamma_{\alpha\beta\beta\alpha}$ increases with coupling strength for fixed energy (see Fig. S1), and the resultant effect is the worsening of Redfield and modified Redfield qualities with increasing coupling strength, especially for small site energy gaps. Note that the participation factor, $\gamma_{\alpha\beta\beta\alpha}$, is a noncontinuous function of coupling strength at zero site energy gap when inhomogeneous disorder is not included (it has a value of 0 for zero coupling and 0.5 otherwise). When inhomogeneous disorder is accounted for realistically at small site energy gaps, however, $\gamma_{\alpha\beta\beta\alpha}$ is a smooth function of the coupling strength, and the phenomenological inclusion of dynamic localization in the theory\cite{Muh2014, Jassas2018} is therefore not necessary to counteract excessive delocalization. Based on the (already) good agreement between the secular ctR and nonsecular FCE spectra at small site energy gaps, such attempts to account for nonsecular effects will likely not improve the quality of Redfield and modified Redfield spectra. Although the exciton gap is largest for the  largest values of the site energy gap and coupling in Fig. \ref{fig:energy coupling grid absorption 300K}, the excitons for these parameters are strongly localized (as seen from the small values for $\gamma_{\alpha\beta\beta\alpha}$ in Fig. S1) and the population transfer rates between excitons are therefore small. Fig. \ref{fig:energy coupling grid absorption 300K}\textbf{a} shows that the qualities of Redfield and modified Redfield spectra improve markedly with the magnitude of the site energy gap, due to the diminishing contribution from population transfer rates. In contrast to the results of Gelzinis \textit{et al.}, the modified Redfield method performs slightly better than the Redfield method for the parameters considered in Fig. \ref{fig:energy coupling grid absorption 300K}.

The qualities in Fig. \ref{fig:energy coupling grid absorption 300K} are generally symmetric in the coupling between sites, but some asymmetries are apparent, the most noticeable of which is the asymmetry between the coupling--energy gap pairs $(55, 220)\pcm$ and $(-55, 220)\pcm$ for the LD qualities in Fig. \ref{fig:energy coupling grid absorption 300K}\textbf{d}. In Fig. \ref{fig:absorption spectra 300K}\textbf{d}, spectra from all the methods are shown for the parameters $(55, 220)\pcm$, and the PI, FCE and ctR spectra are shown for the parameters $(-55, 220)\pcm$. Although the exciton energies, pigment participations, $\bigl(c^{n\alpha}\bigr)^2$ (but not $c^{n\alpha}$), and dipole factor are the same for these two coupling--energy gap pairs, the spectra differ significantly. As discussed in Section \ref{Calculations}, the dipole factor, $f^{1,-1}_{-1,0}$, used in Fig. \ref{fig:energy coupling grid absorption 300K}\textbf{d} corresponds to an LD measurement of a dimer in which the high-energy pigment is orthogonal to the LD axis and the other pigment is at the magic angle with the LD axis. For $(-55, 220)\pcm$, the high-energy exciton contributes most intensely to the spectrum. For $(55, 220)\pcm$, however, the spectral contribution is most intense for the exciton to which the LD-forbidden low-energy pigment contributes predominantly. This borrowing of excitation directly illustrates excitonic quantum superposition between states of local excitation and shows the principle by which forbidden transitions can be probed through their coupling with allowed transitions. Notice that the secular spectra differ significantly from PI and FCE spectra for $(55, 220)\pcm$, which indicates that nonsecular transfer effects are important when excitons cannot be excited directly through interaction with light.

The dependence of the spectral quality on the site energy gap and coupling at $100 \text{ K}$ is qualitatively very similar to the dependence shown in Fig. \ref{fig:energy coupling grid absorption 300K}. An image plot of spectral qualities at $100 \text{ K}$, similar to Fig. \ref{fig:energy coupling grid absorption 300K}, is shown in Fig. S3.

\subsection{Dependence of spectral quality on the excitonic energy gap}
\label{Dependence of spectral quality on the excitonic energy gap}
\begin{figure}[!htbp]
\includegraphics[scale=1, trim=0 1.5cm 0 2cm, clip]{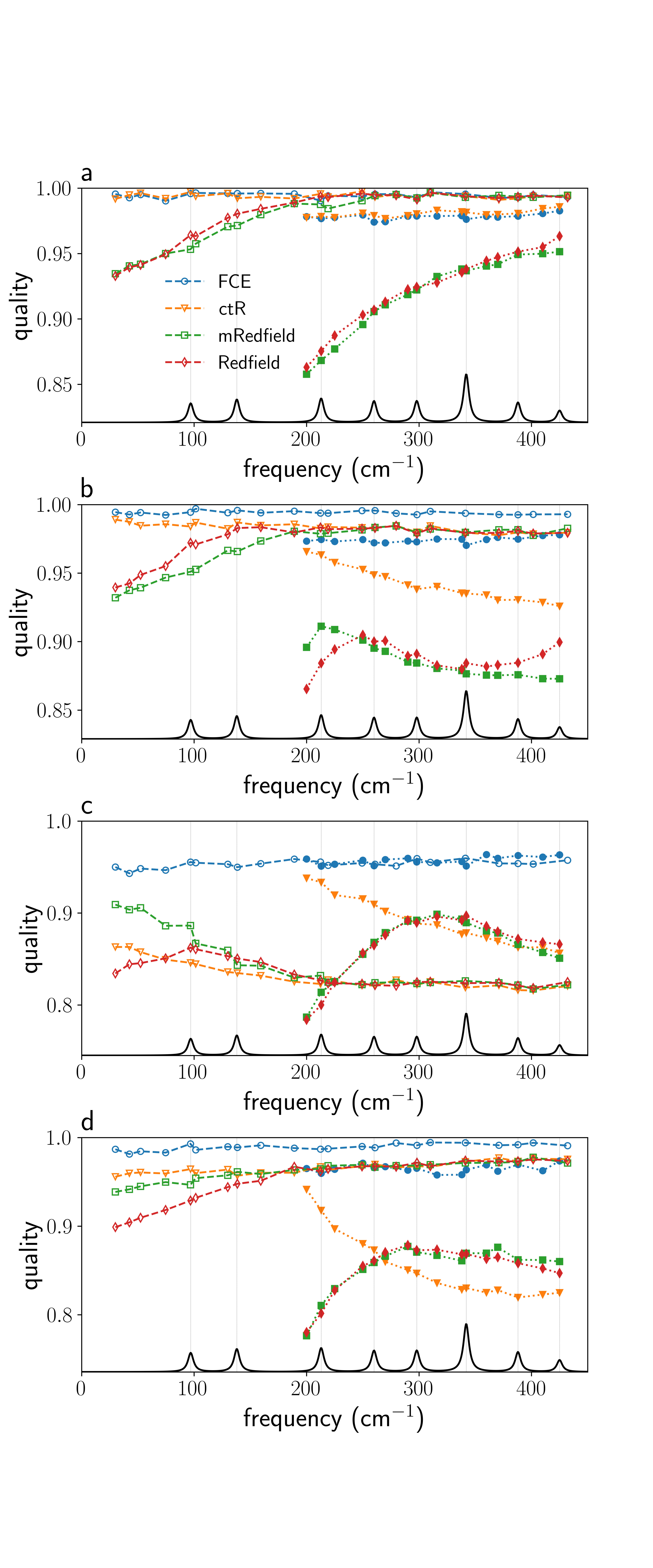}%
\caption{\label{fig:energy varying absorption 300K disorder 140} Quality of absorption-type spectra at 300 K for excitonic coupling $J=15\pcm$ (open markers) and $J=100\pcm$ (filled markers) as a function of the excitonic energy gap.  Qualities are shown for the absorption dipole factors (\textbf{a}) $f^{1,0}_{0,1}$ and (\textbf{b}) $f^{1,1}_{1,1}$, (\textbf{c}) CD dipole factor  $f^{0,1}_{1,0}$, and (\textbf{d}) LD dipole factor  $f^{1,-1}_{-1,0}$. A disorder of $\sigma_{\text{FWHM}}=140\pcm$ was used for approximate spectra. Note that the smallest possible exciton gap equals $2J$. The peak heights of the high-frequency spectral density (black line) are proportional to the Huang-Rhys factors of the intramolecular modes.}
\end{figure}

For the calculations in this section, the excitonic coupling was set to $J=15\pcm$ (an average value for most photosynthetic light-harvesting complexes)  or $J=100\pcm$ (a strong coupling for these complexes), and the site energy gap $\epsilon$ was varied from $0$ to $500\pcm$. For each site energy gap, the excitonic energy gap was calculated as the difference between exciton energies. Fig. \ref{fig:energy varying absorption 300K disorder 140} shows the quality of inhomogeneously broadened approximate absorption-type spectra as a function of the excitonic energy gap at $300 \text{ K}$. The high-frequency spectral density is also included in these figures and was multiplied by a factor $1/\omega^2$ so that the peak heights of the underdamped modes are proportional to their Huang-Rhys factors. As discussed in Section \ref{Calculations}, these underdamped modes correspond to the first eight (from a total of 48) high-frequency vibrations of LHCII, which are mostly  due to intramolecular vibrations in the chlorophyll \textit{a} molecule. The sixth vibrational mode has the largest Huang-Rhys factor of all the modes of LHCII, and the spectral density considered in this article is therefore a good model of the system-bath coupling in a complex system like LHCII. 

It is clear from Fig. \ref{fig:energy varying absorption 300K disorder 140} that, at $300 \text{ K}$, the FCE method produces accurate spectra for moderate ($J=15\pcm$) and strong ($J=100\pcm$) coupling for all of the considered dipole factors. This method performs worst for the calculation of CD spectra, for which it is still $95\%$ accurate. As discussed in Section \ref{Dependence of spectral quality on site energy and excitonic coupling}, the full cumulant expansion is inexact due to the contribution of excitonic propagators to the absorption tensor. For a given exciton gap, the excitation delocalization scales with the coupling strength, and the pigment participation factors $\gamma_{\alpha\beta\beta\alpha}$ (with $\alpha\neq\beta$) therefore increase with coupling (see Fig. S1). Since $\gamma_{\alpha\beta\beta\alpha}$ are the coefficients of terms in Eq. (\ref{FCE K}) that contain excitonic propagators, the FCE method is less accurate for strong coupling than for moderate coupling (for the same exciton gap). As noted in Section \ref{Dependence of spectral quality on site energy and excitonic coupling}, the near independence of FCE qualities on Hamiltonian parameters allows this method to be used for the calculation of absorption-type spectra with a constant, predictable error.

For moderate coupling, the ctR absorption and LD spectra of which the qualities are shown in Figs. \ref{fig:energy varying absorption 300K disorder 140}\textbf{a}, \textbf{b}, and \textbf{d}, deviate less than $4\%$ from the exact spectra, and the quality of these spectra is also nearly constant over the range of exciton energy gaps considered. For strong coupling, the quality of ctR absorption spectra corresponds with the quality of FCE spectra when the secular approximation is valid (Fig. \ref{fig:energy varying absorption 300K disorder 140}\textbf{a}), but deviates significantly from the exact spectra for other dipole factors (Fig. \ref{fig:energy varying absorption 300K disorder 140}\textbf{b-d}). The CD spectra are produced by dynamics of the electronic coherence between the ground state and \textit{delocalized} excited states. The more delocalized the excitons are, the smaller the relative contribution of nonsecular dynamics to the CD spectrum, and the more accurately the spectrum can be calculated with a secular method. For this reason, the ctR CD spectra have a higher quality for small energy gaps or for strong coupling, when excitons are more delocalized, than for large gaps or moderate coupling. As discussed in Section \ref{Dependence of spectral quality on site energy and excitonic coupling}, the electronic transition on the second pigment is LD-forbidden for the dipole factor $f^{1,-1}_{-1,0}$, and excitons that are predominantly localized on this pigment contribute to the LD spectrum mainly via transfer from other excitons. When excitons are fully delocalized, such transfer occurs due to population relaxation, as can be seen by comparing the coefficients for population transfer $\gamma_{\alpha\beta\beta\alpha}$ and nonsecular transfer $\gamma_{\alpha\beta\beta\beta}$ in Fig. S1 for low site energy gaps. For strong coupling and larger energy gaps, however, $\gamma_{\alpha\beta\beta\beta}$ is comparable to or greater than $\gamma_{\alpha\beta\beta\alpha}$ (see Fig. S1), indicating that nonsecular transfer effects are important for the accurate description of LD spectra that depend on interexciton dynamics.

For moderate coupling, the quality of Redfield and modified Redfield spectra at excitonic energy gaps larger than about $250 \pcm$ corresponds almost exactly with the quality of ctR spectra. As discussed in Section \ref{Dependence of spectral quality on site energy and excitonic coupling}, the excitons are localized at large energy gaps and differences between the rate calculations for these methods have a negligible effect. For strong coupling, the difference in accuracy between ctR spectra and Redfield-type spectra also decreases with exciton gap, but is still present for large gaps in Fig. \ref{fig:energy varying absorption 300K disorder 140}, since excitons are not yet fully localized when the coupling is strong. The inaccuracy in the calculation of Redfield and modified Redfield spectra may lead to their quality being incidentally better than for ctR spectra (Fig. \ref{fig:energy varying absorption 300K disorder 140}\textbf{c, d}).

The quality of absorption-type spectra at $100 \text{ K}$ (Fig. S4) depends similarly on the excitonic energy gap as the quality at $300 \text{ K}$, although at $100 \text{ K}$, the Redfield and modified Redfield methods produce even poorer qualities for small exciton energy gaps than they do at $300 \text{ K}$. 

The quality of absorption-type spectra with zero disorder is shown in Figs. S5 (for $T=100 \text{ K}$ and $J=100\pcm$) and S6 (for $T=300 \text{ K}$ and $J=15\pcm$). For each coupling strength, the overall accuracy of these spectra depends similarly on the excitonic energy gap as the accuracy for inhomogeneously broadened spectra. For zero disorder, however, spectra calculated with the Redfield method are significantly less accurate than inhomogeneously broadened spectra when the energy gap is resonant with intramolecular modes. No decrease in accuracy around resonance frequencies is seen for any of the other methods in Figs. S5 and S6. For the calculation of modified Redfield rates, the Fourier-transformed correlation function is convoluted with the Fourier transform of several other time-dependent functions (see Eq. (\ref{modified Redfield rate})), and the accuracy therefore depends more smoothly on the frequency. The sensitivity of Redfield spectra to resonance agrees well with the results obtained by Gelzinis \textit{et al.}\cite{Gelzinis2015}, where the authors used a single intramolecular mode with reorganization energy $\lambda=15\pcm$, comparable to the reorganization energy of $\approx 17\pcm$ of the sixth mode in our study. The insensitivity of modified Redfield qualities to resonance is in contrast to the small but significant decrease in quality seen by Gelzinis \textit{et al.} around the resonance frequency in their study. When using a single-mode spectral density with the parameters of Gelzinis \textit{et al.}, the quality of modified Redfield spectra was indeed lowest at resonance (not shown) but was less sensitive to the excitonic energy gap than the quality of Gelzinis \textit{et al.}, likely due to the different forms used for the underdamped modes in our study and theirs (compare Eq. (\ref{underdamped spectral density}) with Eq. (31) in their article\cite{Gelzinis2015}). For the full spectral density used in our study, the insensitivity of the modified Redfield quality to resonance with a particular underdamped mode may be attributed to the convoluting influence of the other modes.

As discussed in Section \ref{Standard and modified Redfield theories}, the Redfield and modified Redfield methods depend on the long-time rates---the integrands in Eqs. (\ref{Redfield rate}) and (\ref{modified Redfield rate}) may persist for several picoseconds, especially at low temperature. For the temperatures considered in this study, however, the absorption tensor decays in a few hundred femtoseconds and the short-time dynamics of population transfer are therefore important. The necessity of accurately accounting for the short-time dynamics is illustrated by the fact that the quality of spectra calculated using the complex long-time ctR rates (see Fig. S2) is also strongly sensitive to resonance with intramolecular modes (not shown). The non-Markovianity of methods, and not merely the presence of imaginary contributions, therefore helps to prevent sensitivity of spectra to resonance with intramolecular modes. Apart from the non-complexity of their rate contributions (as discussed above), the Redfield and modified Redfield methods are also impacted by the Markovianity of their rates, and these methods therefore perform significantly worse than ctR and FCE.

The sensitivity of quality to resonance is significantly reduced at 300 K and $J=15\pcm$ (Fig. S6), when the weaker coupling causes localization already for small exciton gaps. When inhomogeneous disorder is treated realistically, the accuracy of spectra is independent of resonance with intramolecular modes and is therefore of no concern for the calculation of disordered bulk absorption-type spectra.

As discussed in Section \ref{Dependence of spectral quality on site energy and excitonic coupling}, correlation functions of the form $C_{\alpha\beta\beta\beta}$ in Eq. (\ref{FCE K}) are zero for the site energy gap being zero (corresponding to the smallest exciton gap) and nonsecular effects are therefore absent. For this reason, the ctR and FCE methods correspond exactly for the smallest exciton gap when the disorder is zero, and compare well when disorder is included.

The ratio of qualities obtained for the full spectral density and for a spectral density in which only the fourth intramolecular mode was included, is shown in Fig. S9 for $J=100\pcm$ at $300\text{ K}$. This ratio does not depend straightforwardly on the excitonic energy gap, but is close to one for FCE and ctR spectra, indicating that the accuracy of these methods is independent of the number of high-frequency modes, and the discussion in this article pertaining to the FCE and ctR methods is likely valid for the full spectral density (with 48 modes). The quality of spectra calculated using the Redfield and modified Redfield methods may be significantly poorer for the spectral density with eight modes than for the one with a single mode, and spectra calculated with the full (48 mode) spectral density may therefore be of poorer quality than what is predicted by the analysis in this study. 

\subsection{Dependence of spectral quality on the dipole factor}
\label{Dependence of quality on dipole factor}
In Sections \ref{Dependence of spectral quality on site energy and excitonic coupling} and \ref{Dependence of spectral quality on the excitonic energy gap}, qualities and spectra were shown for four dipole factor matrices. In this section, we determine the dependence of the quality of absorption-type spectra on a generalized dipole factor.

\begin{figure*}[!htbp]
\includegraphics[scale=1]{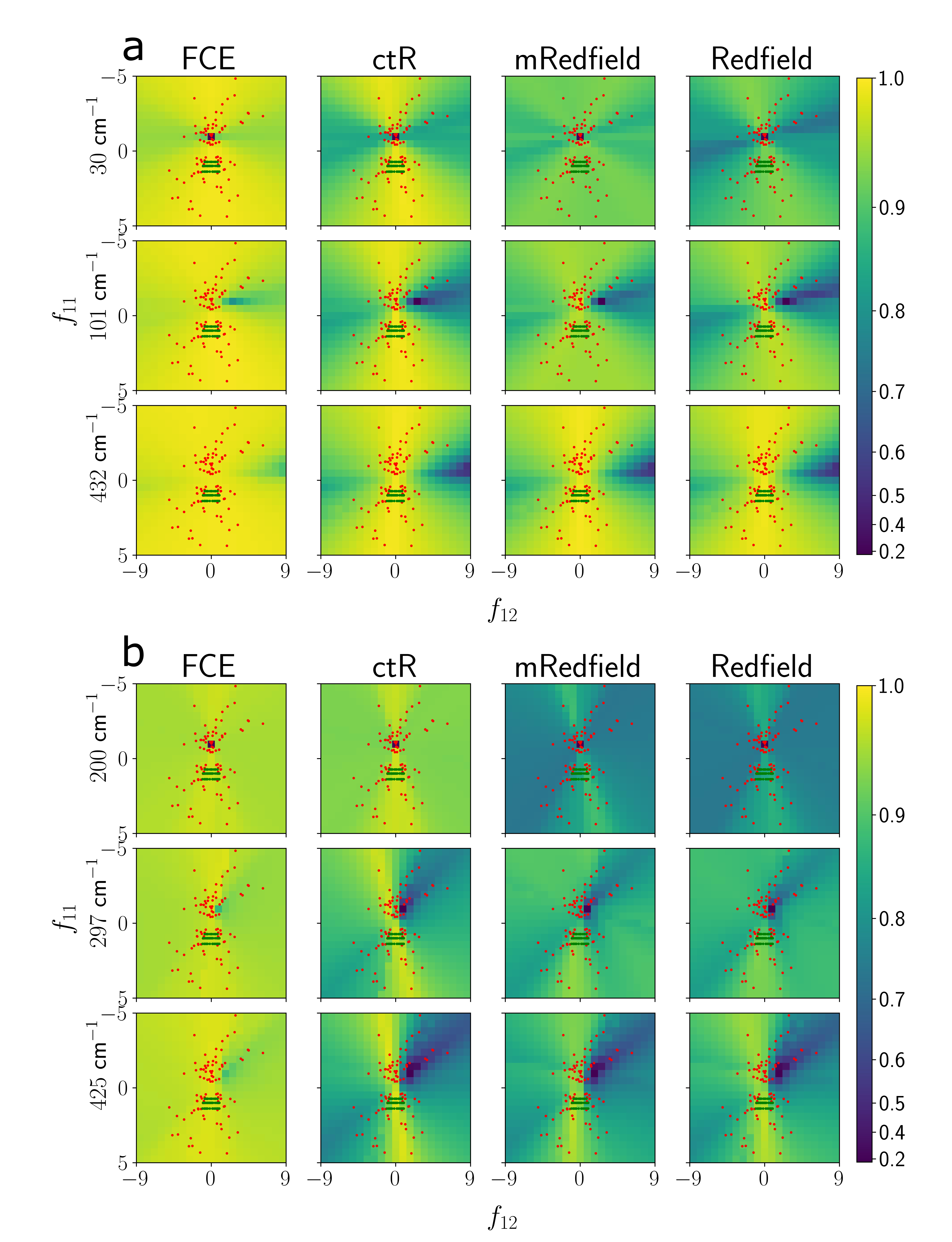}%
\caption{\label{fig:quality as a function of fmn 300 K absorption} Quality of absorption-type spectra at $300 \text{ K}$ as a function of the dipole factor $f^{f_{11},f_{12}}_{f_{12},\,1}$ at the excitation energy gaps indicated for (\textbf{a}) $J=15\pcm$ and (\textbf{b}) $J=100\pcm$. The other parameters are the same as for the calculations for Fig. \ref{fig:energy varying absorption 300K disorder 140}. The dipole factors for LD and OD spectra of CP29 are indicated by red and green dots, respectively, and were calculated from the PDB structure 5xnl.pdb using a dipole strength of $13.96 \text{ D}^2$ for Chl \textit{a} as in Ref. \onlinecite{Mascoli2020}.}
\end{figure*}

To calculate the quality of approximate spectra, we normalized the spectra to their absolute maxima. Only the ratios of the dipole factor matrix elements in Eq. (\ref{Absorption spectrum}) are therefore important. Figs. \ref{fig:quality as a function of fmn 300 K absorption}\textbf{a} and \textbf{b} show the quality of absorption-type spectra for $J=15\pcm$  and $J=100\pcm$, respectively, at $300 \text{ K}$ as a function of the off-diagonal and diagonal element of the (bottom-right normalized) dipole factor matrix
\begin{equation}
   \label{scaled dipole factor matrix}
   f^{f_{11},f_{12}}_{f_{12},\,1} = \begin{bmatrix}
   f_{11} & f_{12}\\
   f_{12} & 1
   \end{bmatrix}.
\end{equation}

This normalization can only be performed when the bottom right element of the dipole factor matrix is nonzero. For this reason, the dipole factor for CD, $f^{0,1}_{1,0}$, and the dipole factor for LD---with the dipole moment of the second pigment at the magic angle relative to the LD axis---$f^{1,-1}_{-1,0}$, are both mapped to $f_{12}\to\pm\infty$.

The assessments of the accuracy for the different methods made in Sections \ref{Dependence of spectral quality on site energy and excitonic coupling} and \ref{Dependence of spectral quality on the excitonic energy gap} are evident in Fig. \ref{fig:quality as a function of fmn 300 K absorption}. The quality of FCE spectra is high for all dipole factors but is best when the magnitude of the diagonal element of the dipole factor is large compared to that of the off-diagonal elements. For moderate coupling (Fig. \ref{fig:quality as a function of fmn 300 K absorption}\textbf{a}), the quality of ctR spectra is also high for dipole factors in the same ranges, but it is significantly worse than the quality of FCE spectra when the magnitudes of the off-diagonal elements of the dipole factor are large. For strong coupling, dipole factors with large off-diagonal magnitudes produce more accurate ctR spectra than they do for moderate coupling. The reasons for the last two observations were discussed in Section \ref{Dependence of spectral quality on the excitonic energy gap} for CD spectra, for which the off-diagonal elements are infinitely large compared to the diagonal elements. The ctR method performs as good as FCE for the dipole factor $f^{1,0}_{0,1}$, but its accuracy (and hence the accuracy of the secular approximation) for other dipole factors depends on the coupling strength. For moderate coupling, nonsecular transfer is weaker than for strong coupling, and the secular approximation is valid for a broader range of off-diagonal dipole factor elements. When the coupling is strong, the spectral quality for some LD dipole factors is worse than for CD dipole factors, and the quality depends nontrivially on the dipole factor. The accuracy of the secular approximation, as a function of the dipole factor, can be modeled straightforwardly by considering the transformation of a  time-independent matrix that summarizes the absorption tensor. Such a model is shown in Fig. S7 for the exciton gaps $101\pcm$ ($J=15\pcm$) and $297\pcm$ ($J=100\pcm$) and corresponds well with the ctR qualities in Fig. \ref{fig:quality as a function of fmn 300 K absorption}. Notice that the FCE qualities show a similar dependence on the dipole factor as ctR (albeit with less sensitivity) and as the expected accuracy of the secular approximation (Fig. S7). This dependence is explained by noting that the diagonal elements of the exciton-basis absorption tensor contains terms for which the second-order cumulant expansion is exact (Eq. (\ref{lineshape function})) together with inexact terms that contain excitonic propagators (Eq. (\ref{ctR rate term})), whereas the off-diagonal elements only contain terms with  propagators. For spectra with strong relative contribution from the  diagonal tensor elements (i.e., spectra for which the secular approximation performs well), the relative contribution of the terms with excitonic propagators is smallest, and the FCE method therefore performs the best.

The quality of Redfield and modified Redfield spectra depends similarly on the dipole factor as the ctR spectra for the exciton energy gaps shown in Fig. \ref{fig:quality as a function of fmn 300 K absorption}, but these techniques perform worse than ctR for all dipole factors.

At the smallest possible excitonic energy gap ($30\pcm$ for $J=15\pcm$ and $200\pcm$ for $J=100\pcm$), the quality of absorption-type spectra is nearly zero for the dipole factor $(f_{11}, f_{12}) = (-1,0)$. For this dipole factor, the LD spectrum is close to zero and the error in the stochastic integration is large---and the quality factor is therefore meaningless. Note that many of the LD dipole factors of CP29 correspond with $(f_{11}, f_{12}) = (-1,0)$ and therefore may contribute little to the total LD spectrum.

At the energy gap of $101 \pcm$ for $J=15\pcm$ and the dipole factor characterized by $(f_{11}, f_{12}) = (-1,3)$, the FCE and PI spectra contain clear vibronic contributions (as shown in Fig. S8), which are absent in the results of the secular methods. Due to the strong vibronic contribution, the FCE and PI spectra agree qualitatively well with each other but differ significantly from the other spectra, which is the cause for the poor quality produced by the ctR, Redfield and modified Redfield methods. 

For moderate coupling, almost all of the absorption and LD dipole factors of the light-harvesting complex CP29 lie in the domain for which the FCE and ctR methods are accurate (as determined for the dimer considered in this study). As discussed above, the CD spectra deviate significantly from the exact spectra. Note, however, that contribution from the intrinsic CD strength of pigment molecules\cite{Lindorfer2017} may cause significant differences between experimental spectra and exact spectra calculated from the excitonic contribution alone. For this reason, the use of accurate methods for the calculation of CD spectra may not be worth their cost. For strong coupling, LD contributions from many of the pigment pairs in CP29 may be considerably impacted by the poor performance of the secular methods. Given the similar dipole strengths of the chlorophyll pigments in other light-harvesting complexes, and the fact that the dipole vectors in CP29 do not have a particular ordered arrangement, the secular LD spectra of light-harvesting complexes that contain strongly coupled pigments may, in general, correspond poorly to exact spectra.

In photosynthetic light-harvesting complexes, the average delocalization length varies from slightly more than one pigment for the PCE545 complex of cryptophytes\cite{Novod_PCE545} to seven or eight pigments in the circularly symmetrical bacterial complexes LH1 and LH2\cite{VanGrondelle2006} and the chlorosome antenna of green sulphur bacteria\cite{D1CP03413H}. The delocalization length in CP29 is estimated to be two to four pigments\cite{Muh2014}. The qualities at the smallest and largest energy gaps in either Fig. \ref{fig:quality as a function of fmn 300 K absorption}\textbf{a} or \textbf{b} represent the cases for delocalized and predominantly localized excitons in a dimer, respectively. Based on the similarity between the dependence of these qualities on the dipole factor for FCE, we conclude that the results of this study will likely extend to aggregates with multiple pigments and that the FCE method may therefore be applied to calculate accurate absorption-type spectra of aggregates with multiple pigments, irrespective of the coupling between pigments. The same reasoning may be applied to the ctR method, and absorption and LD spectra may therefore likely be calculated qualitatively accurately with the ctR method when all the pigments in a pigment aggregate are moderately coupled (i.e., with couplings on the order of $15 \pcm$). For strong coupling, the spectral quality of ctR differs significantly in its dependence on the dipole factor for localized and delocalized excitons, and extension of results in this study to the case for multiple pigments is not straightforward. Based on the dependence of accuracy on the dipole factor for intermediate delocalized states ($297 \pcm$), however, LD spectra for partially localized states in multipigment aggregates may have poor quality.

\section{Conclusion}
In this study, we assessed the accuracy of approximate linear absorption-type spectra calculated with the FCE, ctR, Redfield and modified Redfield methods. We used a pigment dimer as a model system and considered a realistic model spectral density for photosynthetic light-harvesting complexes in plants. We included inhomogeneous disorder and calculated the accuracy of spectra at $300 \text{ K}$ and  $100 \text{ K}$.
Among the approximate methods considered in this study, FCE performs best for the calculation of all types of linear spectra for all molecular parameters and at all temperatures. This method is most accurate for absorption spectra and least accurate for CD spectra. However, for the latter it is still about 95\% accurate. 
The ctR method performs well for the calculation of absorption and LD spectra when the interpigment coupling is moderately large ($\sim 15\pcm$) but performs poorly for the calculation of CD spectra for moderate coupling. Spectra calculated with the ctR method are generally of much lower quality when strong interpigment coupling is present, but absorption spectra are still at least 93\% accurate (for the parameters in this study). When strong coupling is present, ctR LD spectra may be of lower quality than CD spectra, and the spectra calculated with both of these techniques may differ more than 15\% from the exact spectra.
The Redfield and modified Redfield methods perform worse than the ctR method under nearly all circumstances and most notably when the coupling between pigment molecules is strong or the difference between site energies is small, and should not be used if a more accurate method is available.
The quality of approximate spectra is not sensitive to resonance of the exciton gap with intramolecular modes when inhomogeneous disorder is included realistically, and the spectral accuracy is nearly independent of the temperature.

Future work may focus on quantitatively extending the results in this study to pigment aggregates with more than two pigment molecules, especially when strong interpigment coupling is present and on determining the dependence of spectral quality for fluorescence spectra calculated with these methods.

\section*{Acknowledgement}
We acknowledge the use of the University of Pretoria Physics Department's SuperMicro Power-SERVER in this study.
J.A.N. acknowledges financial support from the South African National Research Foundation (grant number 101404) and South African Quantum Technology Initiative (grant number SAQuTI03/2021).
T.P.J.K. acknowledges funding from the National Laser Centre Rental Pool Programme (grant number LRERA13).
T.M. thanks the Czech Science Foundation (GA\v{C}R) Grant number 22-26376S  for financial support.

\section*{Data availability statement}
Data available on reasonable request from the authors.

\appendix
\section{Dipole factor for linear dichroism of disk-shaped complexes}
\label{Dipole factor for linear dichroism of disk-shaped complexes}
An LD spectrum is the difference between absorption spectra measured for two light beams with mutually perpendicular polarization vectors. For a beam propagating in the (Cartesian) $\bm{\hat{y}}$ direction, the LD spectrum is
\begin{equation}
    \label{LD spectrum}
    \text{I}^{\text{LD}}(\omega) = \text{I}_z^{\text{A}}(\omega) - \text{I}_x^{\text{A}}(\omega),
\end{equation}
where $x$ and $z$ indicates the polarization of the light used to illuminate the sample.
LD spectra of small molecules are often measured by means of gel compression. For this technique, complexes are dissolved in a polyacrylamide solvent, which forms a gel\cite{Zucchelli1994}. The gel is then compressed along the $\bm{\hat{x}}$ and $\bm{\hat{y}}$ dimensions by a factor $k \in (0,1)$ and allowed to expand along the $\bm{\hat{z}}$ dimension. The LD spectrum (Eq. (\ref{LD spectrum})) is measured in the compressed state.

To derive the dipole factor $ f^{\mu, \text{LD}}_{mn}$ in Eq. (\ref{LD dipole factor}), we perform the following mental experiment. We consider a small macroscopic volume $V$ of the uncompressed gel containing many disk-shaped complexes, each with $N$ pigments. The complexes are randomly oriented and therefore produce no LD signal. Without changing their orientations, we take all the complexes in $V$ and arrange them on the surface of a sphere that fits within $V$, such that the normal vectors to the discs (which we assume can be assigned unambiguously) point radially outward. Since we measure the far-field LD spectrum, no signal is produced from this arrangement. We then compress the gel. The complexes now lie on the surface of an ellipsoid of which we can determine the geometry from the compression factor $k$. For this arrangement, the LD signal is nonzero. We now place the complexes at their original positions on the surface of the sphere, again without changing their orientations. While the complexes are spaced uniformly on the surface of the sphere, their normal vectors do not point radially outward. Instead, in spherical coordinates, the zenith angle $\beta$ for the normal vectors can be expressed in terms of the zenith angle $\phi$ of their position vectors (which do point radially outward) as
\begin{equation}
    \label{zenith angle of LD normal}
    \beta = \frac{\pi}{2} - \tan^{-1}(k \tan(\frac{\pi}{2} - \phi)).
\end{equation}
Now consider the dipole moment of pigment $m$, $\bm{\mu}_m$, that makes an angle $\alpha$ with the normal vector of its disc. Since the molecules were randomly oriented originally, the many dipoles $\bm{\mu}_m$ at any coarse-grained position on the spherical surface, populate a cone. For the complexes on the pole ($\phi=\beta=0$), the dipole cone is parameterized as:
\begin{equation}
    \label{cone for complexes on the pole}
    \begin{bmatrix}
        \mu_{m,x}\\
        \mu_{m,y}\\
        \mu_{m,z}\\
    \end{bmatrix}
    =|\bm{\mu}_m|
    \begin{bmatrix}
        \sin\alpha\cos t \\ 
        \sin\alpha\sin t \\
        \cos\alpha.
    \end{bmatrix}
    \;\; t\in(0,2\pi)
\end{equation}

In general, for a disk normal vector pointing in the spherical angular direction $(\beta,\gamma)$, the parameterized dipole cone is determined from Eq. (\ref{cone for complexes on the pole}) by consecutive Euler rotations as
\begin{widetext}
\begin{equation}
    \label{cone for complexes with general normal vector}
    \begin{bmatrix}
        \mu_{m,x}\\
        \mu_{m,y}\\
        \mu_{m,z}\\
    \end{bmatrix}
    =|\bm{\mu}_m|
    \begin{bmatrix}
       \cos\gamma(\cos\beta\sin\alpha\cos t + \sin\beta\cos\alpha) - \sin\gamma\sin\alpha\sin t \\
       \sin\gamma(\cos\beta\sin\alpha\cos t + \sin\beta\cos\alpha) + \cos\gamma\sin\alpha\sin t \\
       -\sin\beta\sin\alpha\cos t + \cos\beta\cos\alpha \\
    \end{bmatrix}.
\end{equation}
\end{widetext}

We can now calculate the LD dipole factor by performing a surface integral over the surface of the sphere (over which the complexes are uniformly distributed). We also integrate over the coordinate $t$ that parameterizes the dipole cone around the disk normal:
\begin{eqnarray}
    f^{\mu, \text{LD}}_{mn} = &\int_0^{2\pi}dt\int_0^{2\pi}d\gamma\int_0^{\pi/2}d\phi\,
    \bigl[\mu_{m,z}\mu_{n,z} & \\ \nonumber
    &- \mu_{m,x}\mu_{n,x}\bigr]\sin\phi.
\end{eqnarray}
By using the parametrization of Eq. (\ref{cone for complexes with general normal vector}) and a result from Eq. (\ref{zenith angle of LD normal}),
\begin{align}
\cos\beta = \frac{k\cot\phi}{\sqrt{k^2\cot^2\phi+1}} && \sin\beta = \frac{1}{\sqrt{k^2\cot^2\phi+1}},
\end{align}
we obtain (after some algebra and online integral calculation):
\begin{eqnarray}
    f^{\mu, \text{LD}}_{mn} =&& \pi^2\bigl(\bm{\mu}_m\cdot\bm{\mu}_n - 3|\bm{\mu}_m||\bm{\mu}_n|\cos\alpha_m\cos\alpha_n\bigr)\nonumber\\
                             &&\times\bigl(3L - 2\bigr),
\end{eqnarray}
with
\begin{eqnarray}
    L &&= \int_0^{\pi/2}d\phi\,\frac{\sin^3\phi}{k^2\cos^2\phi + \sin^2\phi} \nonumber\\
      &&=-\frac{(1-u^2)\ln(\frac{1+u}{1-u}) - 2u}{2u^3},
\end{eqnarray}
where $u = \sqrt{1 - k^2}$.
In general, one is interested only in the relative intensity of the LD spectrum, and we simply define
\begin{equation}
f^{\mu, \text{LD}}_{mn} = \bm{\mu}_m\cdot\bm{\mu}_n - 3|\bm{\mu}_m||\bm{\mu}_n|\cos\alpha_m\cos\alpha_n.
\end{equation}

\section{Second-order weak scheme}
\label{Second order weak scheme}
The explicit order 2 weak scheme that we use to propagate the stochastic matrices can be derived by a straightforward but somewhat tedious calculation from the vector expressions in Chapter 15 of Kloeden and Platen\cite{kloeden2011numerical}:
\begin{eqnarray}
   \rho_i =& \bigl[I + \frac{1}{2}\Delta -  (\Delta H + \Delta \Xi_{i-1}) \nonumber \\ 
   &\cdot(iI + \frac{1}{2}\Delta H + \frac{1}{2}\Delta \Xi_{i-1})\bigr]\rho_{i-1}\label{second order weak scheme},
\end{eqnarray}
where, for a given discretization, $I$ is the identity matrix, $\Delta = \delta tI$, and $\Xi_{i-1}$ is a diagonal matrix with $\xi_{n,i-1}$ on the $n^\text{th}$ diagonal position. In Eq. (\ref{second order weak scheme}), $(\cdot)$ denotes matrix multiplication.

\bibliography{library}

\begin{thebibliography}{37}%
\makeatletter
\providecommand \@ifxundefined [1]{%
 \@ifx{#1\undefined}
}%
\providecommand \@ifnum [1]{%
 \ifnum #1\expandafter \@firstoftwo
 \else \expandafter \@secondoftwo
 \fi
}%
\providecommand \@ifx [1]{%
 \ifx #1\expandafter \@firstoftwo
 \else \expandafter \@secondoftwo
 \fi
}%
\providecommand \natexlab [1]{#1}%
\providecommand \enquote  [1]{``#1''}%
\providecommand \bibnamefont  [1]{#1}%
\providecommand \bibfnamefont [1]{#1}%
\providecommand \citenamefont [1]{#1}%
\providecommand \href@noop [0]{\@secondoftwo}%
\providecommand \href [0]{\begingroup \@sanitize@url \@href}%
\providecommand \@href[1]{\@@startlink{#1}\@@href}%
\providecommand \@@href[1]{\endgroup#1\@@endlink}%
\providecommand \@sanitize@url [0]{\catcode `\\12\catcode `\$12\catcode
  `\&12\catcode `\#12\catcode `\^12\catcode `\_12\catcode `\%12\relax}%
\providecommand \@@startlink[1]{}%
\providecommand \@@endlink[0]{}%
\providecommand \url  [0]{\begingroup\@sanitize@url \@url }%
\providecommand \@url [1]{\endgroup\@href {#1}{\urlprefix }}%
\providecommand \urlprefix  [0]{URL }%
\providecommand \Eprint [0]{\href }%
\providecommand \doibase [0]{http://dx.doi.org/}%
\providecommand \selectlanguage [0]{\@gobble}%
\providecommand \bibinfo  [0]{\@secondoftwo}%
\providecommand \bibfield  [0]{\@secondoftwo}%
\providecommand \translation [1]{[#1]}%
\providecommand \BibitemOpen [0]{}%
\providecommand \bibitemStop [0]{}%
\providecommand \bibitemNoStop [0]{.\EOS\space}%
\providecommand \EOS [0]{\spacefactor3000\relax}%
\providecommand \BibitemShut  [1]{\csname bibitem#1\endcsname}%
\let\auto@bib@innerbib\@empty
\bibitem [{\citenamefont {Garab}\ and\ \citenamefont {van
  Amerongen}(2009)}]{Garab2009}%
  \BibitemOpen
  \bibfield  {author} {\bibinfo {author} {\bibfnamefont {G.}~\bibnamefont
  {Garab}}\ and\ \bibinfo {author} {\bibfnamefont {H.}~\bibnamefont {van
  Amerongen}},\ }\bibfield  {title} {\enquote {\bibinfo {title} {{Linear
  dichroism and circular dichroism in photosynthesis research}},}\ }\href
  {\doibase 10.1007/s11120-009-9424-4} {\bibfield  {journal} {\bibinfo
  {journal} {Photosyn. Res.}\ }\textbf {\bibinfo {volume} {101}},\ \bibinfo
  {pages} {135--146} (\bibinfo {year} {2009})}\BibitemShut {NoStop}%
\bibitem [{\citenamefont {Rodger}\ and\ \citenamefont
  {Nord{\'{e}}n}(1997)}]{rodger1997circular}%
  \BibitemOpen
  \bibfield  {author} {\bibinfo {author} {\bibfnamefont {A.}~\bibnamefont
  {Rodger}}\ and\ \bibinfo {author} {\bibfnamefont {B.}~\bibnamefont
  {Nord{\'{e}}n}},\ }\href {https://books.google.co.za/books?id=THeKGC99hJcC}
  {\emph {\bibinfo {title} {{Circular Dichroism and Linear Dichroism}}}},\
  Oxford Classical Monographs\ (\bibinfo  {publisher} {Oxford University
  Press},\ \bibinfo {year} {1997})\BibitemShut {NoStop}%
\bibitem [{\citenamefont {Romero}\ \emph {et~al.}(2009)\citenamefont {Romero},
  \citenamefont {Mozzo}, \citenamefont {van Stokkum}, \citenamefont {Dekker},
  \citenamefont {van Grondelle},\ and\ \citenamefont {Croce}}]{Romero2009}%
  \BibitemOpen
  \bibfield  {author} {\bibinfo {author} {\bibfnamefont {E.}~\bibnamefont
  {Romero}}, \bibinfo {author} {\bibfnamefont {M.}~\bibnamefont {Mozzo}},
  \bibinfo {author} {\bibfnamefont {I.~H.}\ \bibnamefont {van Stokkum}},
  \bibinfo {author} {\bibfnamefont {J.~P.}\ \bibnamefont {Dekker}}, \bibinfo
  {author} {\bibfnamefont {R.}~\bibnamefont {van Grondelle}}, \ and\ \bibinfo
  {author} {\bibfnamefont {R.}~\bibnamefont {Croce}},\ }\bibfield  {title}
  {\enquote {\bibinfo {title} {{The origin of the low-energy form of
  Photosystem I light-harvesting complex Lhca4: mixing of the lowest exciton
  with a charge-transfer state}},}\ }\href {\doibase 10.1016/j.bpj.2008.11.043}
  {\bibfield  {journal} {\bibinfo  {journal} {Biophys. J.}\ }\textbf {\bibinfo
  {volume} {96}},\ \bibinfo {pages} {L35--L37} (\bibinfo {year}
  {2009})}\BibitemShut {NoStop}%
\bibitem [{\citenamefont {Br{\"{u}}ggemann}\ \emph {et~al.}(2004)\citenamefont
  {Br{\"{u}}ggemann}, \citenamefont {Sznee}, \citenamefont {Novoderezhkin},
  \citenamefont {van Grondelle},\ and\ \citenamefont {May}}]{Sznee2004}%
  \BibitemOpen
  \bibfield  {author} {\bibinfo {author} {\bibfnamefont {B.}~\bibnamefont
  {Br{\"{u}}ggemann}}, \bibinfo {author} {\bibfnamefont {K.}~\bibnamefont
  {Sznee}}, \bibinfo {author} {\bibfnamefont {V.}~\bibnamefont
  {Novoderezhkin}}, \bibinfo {author} {\bibfnamefont {R.}~\bibnamefont {van
  Grondelle}}, \ and\ \bibinfo {author} {\bibfnamefont {V.}~\bibnamefont
  {May}},\ }\bibfield  {title} {\enquote {\bibinfo {title} {{From structure to
  dynamics: modeling exciton dynamics in the photosynthetic antenna PS1}},}\
  }\href {\doibase 10.1021/jp0401473} {\bibfield  {journal} {\bibinfo
  {journal} {J. Phys. Chem. B}\ }\textbf {\bibinfo {volume} {108}},\ \bibinfo
  {pages} {13536--13546} (\bibinfo {year} {2004})}\BibitemShut {NoStop}%
\bibitem [{\citenamefont {Kr{\"{u}}ger}\ \emph {et~al.}(2010)\citenamefont
  {Kr{\"{u}}ger}, \citenamefont {Novoderezhkin}, \citenamefont {Ilioaia},\ and\
  \citenamefont {van Grondelle}}]{Kruger2010}%
  \BibitemOpen
  \bibfield  {author} {\bibinfo {author} {\bibfnamefont {T.~P.~J.}\
  \bibnamefont {Kr{\"{u}}ger}}, \bibinfo {author} {\bibfnamefont {V.~I.}\
  \bibnamefont {Novoderezhkin}}, \bibinfo {author} {\bibfnamefont
  {C.}~\bibnamefont {Ilioaia}}, \ and\ \bibinfo {author} {\bibfnamefont
  {R.}~\bibnamefont {van Grondelle}},\ }\bibfield  {title} {\enquote {\bibinfo
  {title} {{Fluorescence spectral dynamics of single LHCII trimers}},}\ }\href
  {\doibase 10.1016/j.bpj.2010.03.028} {\bibfield  {journal} {\bibinfo
  {journal} {Biophys. J.}\ }\textbf {\bibinfo {volume} {98}},\ \bibinfo {pages}
  {3093--3101} (\bibinfo {year} {2010})}\BibitemShut {NoStop}%
\bibitem [{\citenamefont {Novoderezhkin}\ \emph {et~al.}(2004)\citenamefont
  {Novoderezhkin}, \citenamefont {Palacios}, \citenamefont {van Amerongen},\
  and\ \citenamefont {van Grondelle}}]{Novoderezhkin2004a}%
  \BibitemOpen
  \bibfield  {author} {\bibinfo {author} {\bibfnamefont {V.~I.}\ \bibnamefont
  {Novoderezhkin}}, \bibinfo {author} {\bibfnamefont {M.~A.}\ \bibnamefont
  {Palacios}}, \bibinfo {author} {\bibfnamefont {H.}~\bibnamefont {van
  Amerongen}}, \ and\ \bibinfo {author} {\bibfnamefont {R.}~\bibnamefont {van
  Grondelle}},\ }\bibfield  {title} {\enquote {\bibinfo {title}
  {{Energy-transfer dynamics in the LHCII complex of higher plants: modified
  Redfield approach}},}\ }\href {\doibase 10.1021/jp0496001} {\bibfield
  {journal} {\bibinfo  {journal} {J. Phys. Chem. B}\ }\textbf {\bibinfo
  {volume} {108}},\ \bibinfo {pages} {10363--10375} (\bibinfo {year}
  {2004})}\BibitemShut {NoStop}%
\bibitem [{\citenamefont {Ramanan}\ \emph {et~al.}(2015)\citenamefont
  {Ramanan}, \citenamefont {Gruber}, \citenamefont {Mal{\'{y}}}, \citenamefont
  {Negretti}, \citenamefont {Novoderezhkin}, \citenamefont {Kr{\"{u}}ger},
  \citenamefont {Man{\v{c}}al}, \citenamefont {Croce},\ and\ \citenamefont {van
  Grondelle}}]{Ramanan2015}%
  \BibitemOpen
  \bibfield  {author} {\bibinfo {author} {\bibfnamefont {C.}~\bibnamefont
  {Ramanan}}, \bibinfo {author} {\bibfnamefont {J.~M.}\ \bibnamefont {Gruber}},
  \bibinfo {author} {\bibfnamefont {P.}~\bibnamefont {Mal{\'{y}}}}, \bibinfo
  {author} {\bibfnamefont {M.}~\bibnamefont {Negretti}}, \bibinfo {author}
  {\bibfnamefont {V.}~\bibnamefont {Novoderezhkin}}, \bibinfo {author}
  {\bibfnamefont {T.~P.~J.}\ \bibnamefont {Kr{\"{u}}ger}}, \bibinfo {author}
  {\bibfnamefont {T.}~\bibnamefont {Man{\v{c}}al}}, \bibinfo {author}
  {\bibfnamefont {R.}~\bibnamefont {Croce}}, \ and\ \bibinfo {author}
  {\bibfnamefont {R.}~\bibnamefont {van Grondelle}},\ }\bibfield  {title}
  {\enquote {\bibinfo {title} {{The role of exciton delocalization in the major
  photosynthetic light-harvesting antenna of plants}},}\ }\href {\doibase
  10.1016/j.bpj.2015.01.019} {\bibfield  {journal} {\bibinfo  {journal}
  {Biophys. J.}\ }\textbf {\bibinfo {volume} {108}},\ \bibinfo {pages}
  {1047--1056} (\bibinfo {year} {2015})}\BibitemShut {NoStop}%
\bibitem [{\citenamefont {Palacios}\ \emph {et~al.}(2002)\citenamefont
  {Palacios}, \citenamefont {de~Weerd}, \citenamefont {Ihalainen},
  \citenamefont {van Grondelle},\ and\ \citenamefont {van
  Amerongen}}]{Palacios2002}%
  \BibitemOpen
  \bibfield  {author} {\bibinfo {author} {\bibfnamefont {M.~A.}\ \bibnamefont
  {Palacios}}, \bibinfo {author} {\bibfnamefont {F.~L.}\ \bibnamefont
  {de~Weerd}}, \bibinfo {author} {\bibfnamefont {J.~A.}\ \bibnamefont
  {Ihalainen}}, \bibinfo {author} {\bibfnamefont {R.}~\bibnamefont {van
  Grondelle}}, \ and\ \bibinfo {author} {\bibfnamefont {H.}~\bibnamefont {van
  Amerongen}},\ }\bibfield  {title} {\enquote {\bibinfo {title} {{Superradiance
  and exciton (de)localization in Light-Harvesting Complex II from green
  plants?}}}\ }\href {\doibase 10.1021/jp014078t} {\bibfield  {journal}
  {\bibinfo  {journal} {J. Phys. Chem. B}\ }\textbf {\bibinfo {volume} {106}},\
  \bibinfo {pages} {5782--5787} (\bibinfo {year} {2002})}\BibitemShut {NoStop}%
\bibitem [{\citenamefont {Bulheller}, \citenamefont {Rodger},\ and\
  \citenamefont {Hirst}(2007)}]{Bulheller2007}%
  \BibitemOpen
  \bibfield  {author} {\bibinfo {author} {\bibfnamefont {B.~M.}\ \bibnamefont
  {Bulheller}}, \bibinfo {author} {\bibfnamefont {A.}~\bibnamefont {Rodger}}, \
  and\ \bibinfo {author} {\bibfnamefont {J.~D.}\ \bibnamefont {Hirst}},\
  }\bibfield  {title} {\enquote {\bibinfo {title} {{Circular and linear
  dichroism of proteins}},}\ }\href {\doibase 10.1039/b615870f} {\bibfield
  {journal} {\bibinfo  {journal} {Phys. Chem. Chem. Phys.}\ }\textbf {\bibinfo
  {volume} {9}},\ \bibinfo {pages} {2020--2035} (\bibinfo {year}
  {2007})}\BibitemShut {NoStop}%
\bibitem [{\citenamefont {Croce}\ \emph {et~al.}(2002)\citenamefont {Croce},
  \citenamefont {Morosinotto}, \citenamefont {Castelletti}, \citenamefont
  {Breton},\ and\ \citenamefont {Bassi}}]{Croce2002}%
  \BibitemOpen
  \bibfield  {author} {\bibinfo {author} {\bibfnamefont {R.}~\bibnamefont
  {Croce}}, \bibinfo {author} {\bibfnamefont {T.}~\bibnamefont {Morosinotto}},
  \bibinfo {author} {\bibfnamefont {S.}~\bibnamefont {Castelletti}}, \bibinfo
  {author} {\bibfnamefont {J.}~\bibnamefont {Breton}}, \ and\ \bibinfo {author}
  {\bibfnamefont {R.}~\bibnamefont {Bassi}},\ }\bibfield  {title} {\enquote
  {\bibinfo {title} {{The Lhca antenna complexes of higher plants Photosystem
  I}},}\ }\href {\doibase 10.1016/S0005-2728(02)00304-3} {\bibfield  {journal}
  {\bibinfo  {journal} {Biochim. Biophys. Acta - Bioenerg.}\ }\textbf {\bibinfo
  {volume} {1556}},\ \bibinfo {pages} {29--40} (\bibinfo {year}
  {2002})}\BibitemShut {NoStop}%
\bibitem [{\citenamefont {Tanimura}\ and\ \citenamefont
  {Kubo}(1989)}]{Tanimura1989}%
  \BibitemOpen
  \bibfield  {author} {\bibinfo {author} {\bibfnamefont {Y.}~\bibnamefont
  {Tanimura}}\ and\ \bibinfo {author} {\bibfnamefont {R.}~\bibnamefont
  {Kubo}},\ }\bibfield  {title} {\enquote {\bibinfo {title} {{Time evolution of
  a quantum system in contact with a nearly Gaussian-Markoffian noise bath}},}\
  }\href {\doibase 10.1143/JPSJ.58.101} {\bibfield  {journal} {\bibinfo
  {journal} {J. Phys. Soc. Japan}\ }\textbf {\bibinfo {volume} {58}},\ \bibinfo
  {pages} {101--114} (\bibinfo {year} {1989})}\BibitemShut {NoStop}%
\bibitem [{\citenamefont {Ishizaki}\ and\ \citenamefont
  {Fleming}(2009)}]{Ishizaki2009}%
  \BibitemOpen
  \bibfield  {author} {\bibinfo {author} {\bibfnamefont {A.}~\bibnamefont
  {Ishizaki}}\ and\ \bibinfo {author} {\bibfnamefont {G.~R.}\ \bibnamefont
  {Fleming}},\ }\bibfield  {title} {\enquote {\bibinfo {title} {{Theoretical
  examination of quantum coherence in a photosynthetic system at physiological
  temperature.}}}\ }\href {\doibase 10.1073/pnas.0908989106} {\bibfield
  {journal} {\bibinfo  {journal} {Proc. Natl. Acad. Sci. U.S.A.}\ }\textbf
  {\bibinfo {volume} {106}},\ \bibinfo {pages} {17255--17260} (\bibinfo {year}
  {2009})}\BibitemShut {NoStop}%
\bibitem [{\citenamefont {Kreisbeck}, \citenamefont {Kramer},\ and\
  \citenamefont {Aspuru-Guzik}(2014)}]{Kreisbeck2014}%
  \BibitemOpen
  \bibfield  {author} {\bibinfo {author} {\bibfnamefont {C.}~\bibnamefont
  {Kreisbeck}}, \bibinfo {author} {\bibfnamefont {T.}~\bibnamefont {Kramer}}, \
  and\ \bibinfo {author} {\bibfnamefont {A.}~\bibnamefont {Aspuru-Guzik}},\
  }\bibfield  {title} {\enquote {\bibinfo {title} {{Scalable high-performance
  algorithm for the simulation of exciton dynamics. Application to the
  Light-Harvesting Complex II in the presence of resonant vibrational
  modes}},}\ }\href {\doibase 10.1021/ct500629s} {\bibfield  {journal}
  {\bibinfo  {journal} {J. Chem. Theory Comput.}\ }\textbf {\bibinfo {volume}
  {10}},\ \bibinfo {pages} {4045--4054} (\bibinfo {year} {2014})}\BibitemShut
  {NoStop}%
\bibitem [{\citenamefont {Novoderezhkin}, \citenamefont {Romero},\ and\
  \citenamefont {van Grondelle}(2015)}]{Novoderezhkin2015}%
  \BibitemOpen
  \bibfield  {author} {\bibinfo {author} {\bibfnamefont {V.~I.}\ \bibnamefont
  {Novoderezhkin}}, \bibinfo {author} {\bibfnamefont {E.}~\bibnamefont
  {Romero}}, \ and\ \bibinfo {author} {\bibfnamefont {R.}~\bibnamefont {van
  Grondelle}},\ }\bibfield  {title} {\enquote {\bibinfo {title} {{How
  exciton-vibrational coherences control charge separation in the Photosystem
  II reaction center}},}\ }\href {\doibase 10.1039/C5CP00582E} {\bibfield
  {journal} {\bibinfo  {journal} {Phys. Chem. Chem. Phys.}\ }\textbf {\bibinfo
  {volume} {17}},\ \bibinfo {pages} {30828--30841} (\bibinfo {year}
  {2015})}\BibitemShut {NoStop}%
\bibitem [{\citenamefont {Moix}, \citenamefont {Ma},\ and\ \citenamefont
  {Cao}(2015)}]{Moix2015}%
  \BibitemOpen
  \bibfield  {author} {\bibinfo {author} {\bibfnamefont {J.~M.}\ \bibnamefont
  {Moix}}, \bibinfo {author} {\bibfnamefont {J.}~\bibnamefont {Ma}}, \ and\
  \bibinfo {author} {\bibfnamefont {J.}~\bibnamefont {Cao}},\ }\bibfield
  {title} {\enquote {\bibinfo {title} {{F{\"{o}}rster resonance energy
  transfer, absorption and emission spectra in multichromophoric systems. III.
  Exact stochastic path integral evaluation}},}\ }\href {\doibase
  10.1063/1.4908601} {\bibfield  {journal} {\bibinfo  {journal} {J. Chem.
  Phys.}\ }\textbf {\bibinfo {volume} {142}} (\bibinfo {year} {2015}),\
  10.1063/1.4908601},\ \Eprint {http://arxiv.org/abs/1501.05679}
  {arXiv:1501.05679} \BibitemShut {NoStop}%
\bibitem [{\citenamefont {Renger}\ and\ \citenamefont
  {Marcus}(2002)}]{Renger2002}%
  \BibitemOpen
  \bibfield  {author} {\bibinfo {author} {\bibfnamefont {T.}~\bibnamefont
  {Renger}}\ and\ \bibinfo {author} {\bibfnamefont {R.~A.}\ \bibnamefont
  {Marcus}},\ }\bibfield  {title} {\enquote {\bibinfo {title} {{On the relation
  of protein dynamics and exciton relaxation in pigment-protein complexes: An
  estimation of the spectral density and a theory for the calculation of
  optical spectra}},}\ }\href {\doibase 10.1063/1.1470200} {\bibfield
  {journal} {\bibinfo  {journal} {J. Chem. Phys.}\ }\textbf {\bibinfo {volume}
  {116}},\ \bibinfo {pages} {9997--10019} (\bibinfo {year} {2002})}\BibitemShut
  {NoStop}%
\bibitem [{\citenamefont {Valkunas}, \citenamefont {Abramavicius},\ and\
  \citenamefont {Man{\v{c}}al}(2013)}]{Valkunas2013}%
  \BibitemOpen
  \bibfield  {author} {\bibinfo {author} {\bibfnamefont {L.}~\bibnamefont
  {Valkunas}}, \bibinfo {author} {\bibfnamefont {D.}~\bibnamefont
  {Abramavicius}}, \ and\ \bibinfo {author} {\bibfnamefont {T.}~\bibnamefont
  {Man{\v{c}}al}},\ }\href {\doibase 10.1002/9783527653652} {\emph {\bibinfo
  {title} {{Molecular Excitation Dynamics and Relaxation}}}}\ (\bibinfo
  {publisher} {Wiley-VCH Verlag GmbH {\&} Co. KGaA},\ \bibinfo {address}
  {Weinheim, Germany},\ \bibinfo {year} {2013})\BibitemShut {NoStop}%
\bibitem [{\citenamefont {Ma}\ and\ \citenamefont {Cao}(2015)}]{Ma2015}%
  \BibitemOpen
  \bibfield  {author} {\bibinfo {author} {\bibfnamefont {J.}~\bibnamefont
  {Ma}}\ and\ \bibinfo {author} {\bibfnamefont {J.}~\bibnamefont {Cao}},\
  }\bibfield  {title} {\enquote {\bibinfo {title} {{F{\"{o}}rster resonance
  energy transfer, absorption and emission spectra in multichromophoric
  systems. I. Full cumulant expansions and system-bath entanglement}},}\ }\href
  {\doibase 10.1063/1.4908599} {\bibfield  {journal} {\bibinfo  {journal} {J.
  Chem. Phys.}\ }\textbf {\bibinfo {volume} {142}},\ \bibinfo {pages} {094106}
  (\bibinfo {year} {2015})},\ \Eprint {http://arxiv.org/abs/1501.05679}
  {arXiv:1501.05679} \BibitemShut {NoStop}%
\bibitem [{\citenamefont {Gelzinis}, \citenamefont {Abramavicius},\ and\
  \citenamefont {Valkunas}(2015)}]{Gelzinis2015}%
  \BibitemOpen
  \bibfield  {author} {\bibinfo {author} {\bibfnamefont {A.}~\bibnamefont
  {Gelzinis}}, \bibinfo {author} {\bibfnamefont {D.}~\bibnamefont
  {Abramavicius}}, \ and\ \bibinfo {author} {\bibfnamefont {L.}~\bibnamefont
  {Valkunas}},\ }\bibfield  {title} {\enquote {\bibinfo {title} {{Absorption
  lineshapes of molecular aggregates revisited}},}\ }\href {\doibase
  10.1063/1.4918343} {\bibfield  {journal} {\bibinfo  {journal} {J. Chem.
  Phys.}\ }\textbf {\bibinfo {volume} {142}} (\bibinfo {year} {2015}),\
  10.1063/1.4918343}\BibitemShut {NoStop}%
\bibitem [{\citenamefont {Jassas}\ \emph {et~al.}(2018)\citenamefont {Jassas},
  \citenamefont {Chen}, \citenamefont {Khmelnitskiy}, \citenamefont {Casazza},
  \citenamefont {Santabarbara},\ and\ \citenamefont {Jankowiak}}]{Jassas2018}%
  \BibitemOpen
  \bibfield  {author} {\bibinfo {author} {\bibfnamefont {M.}~\bibnamefont
  {Jassas}}, \bibinfo {author} {\bibfnamefont {J.}~\bibnamefont {Chen}},
  \bibinfo {author} {\bibfnamefont {A.}~\bibnamefont {Khmelnitskiy}}, \bibinfo
  {author} {\bibfnamefont {A.~P.}\ \bibnamefont {Casazza}}, \bibinfo {author}
  {\bibfnamefont {S.}~\bibnamefont {Santabarbara}}, \ and\ \bibinfo {author}
  {\bibfnamefont {R.}~\bibnamefont {Jankowiak}},\ }\bibfield  {title} {\enquote
  {\bibinfo {title} {{Structure-based exciton Hamiltonian and dynamics for the
  reconstituted wild-type CP29 protein antenna complex of the Photosystem
  II}},}\ }\href {\doibase 10.1021/acs.jpcb.8b00032} {\bibfield  {journal}
  {\bibinfo  {journal} {J. Phys. Chem. B}\ }\textbf {\bibinfo {volume} {122}},\
  \bibinfo {pages} {4611--4624} (\bibinfo {year} {2018})}\BibitemShut {NoStop}%
\bibitem [{\citenamefont {Cupellini}, \citenamefont {Lipparini},\ and\
  \citenamefont {Cao}(2020)}]{Cupellini2020}%
  \BibitemOpen
  \bibfield  {author} {\bibinfo {author} {\bibfnamefont {L.}~\bibnamefont
  {Cupellini}}, \bibinfo {author} {\bibfnamefont {F.}~\bibnamefont
  {Lipparini}}, \ and\ \bibinfo {author} {\bibfnamefont {J.}~\bibnamefont
  {Cao}},\ }\bibfield  {title} {\enquote {\bibinfo {title} {{Absorption and
  circular dichroism spectra of molecular aggregates with the Full Cumulant
  Expansion}},}\ }\href {\doibase 10.1021/acs.jpcb.0c05180} {\bibfield
  {journal} {\bibinfo  {journal} {J. Phys. Chem. B}\ }\textbf {\bibinfo
  {volume} {124}},\ \bibinfo {pages} {8610--8617} (\bibinfo {year}
  {2020})}\BibitemShut {NoStop}%
\bibitem [{\citenamefont {Schr{\"{o}}der}, \citenamefont
  {Kleinekath{\"{o}}fer},\ and\ \citenamefont
  {Schreiber}(2006)}]{Schroder2006}%
  \BibitemOpen
  \bibfield  {author} {\bibinfo {author} {\bibfnamefont {M.}~\bibnamefont
  {Schr{\"{o}}der}}, \bibinfo {author} {\bibfnamefont {U.}~\bibnamefont
  {Kleinekath{\"{o}}fer}}, \ and\ \bibinfo {author} {\bibfnamefont
  {M.}~\bibnamefont {Schreiber}},\ }\bibfield  {title} {\enquote {\bibinfo
  {title} {{Calculation of absorption spectra for light-harvesting systems
  using non-Markovian approaches as well as modified Redfield theory}},}\
  }\href {\doibase 10.1063/1.2171188} {\bibfield  {journal} {\bibinfo
  {journal} {J. Chem. Phys.}\ }\textbf {\bibinfo {volume} {124}},\ \bibinfo
  {pages} {084903} (\bibinfo {year} {2006})}\BibitemShut {NoStop}%
\bibitem [{\citenamefont {Mukamel}(1995)}]{Mukamel1995}%
  \BibitemOpen
  \bibfield  {author} {\bibinfo {author} {\bibfnamefont {S.}~\bibnamefont
  {Mukamel}},\ }\href
  {https://books.google.com/books?id=k{\_}7uAAAAMAAJ{\&}pgis=1} {\emph
  {\bibinfo {title} {{Principles of Nonlinear Optical Spectroscopy}}}}\
  (\bibinfo  {publisher} {Oxford University Press},\ \bibinfo {year}
  {1995})\BibitemShut {NoStop}%
\bibitem [{\citenamefont {Zhang}\ \emph {et~al.}(1998)\citenamefont {Zhang},
  \citenamefont {Meier}, \citenamefont {Chernyak},\ and\ \citenamefont
  {Mukamel}}]{Zhang1998}%
  \BibitemOpen
  \bibfield  {author} {\bibinfo {author} {\bibfnamefont {W.~M.}\ \bibnamefont
  {Zhang}}, \bibinfo {author} {\bibfnamefont {T.}~\bibnamefont {Meier}},
  \bibinfo {author} {\bibfnamefont {V.}~\bibnamefont {Chernyak}}, \ and\
  \bibinfo {author} {\bibfnamefont {S.}~\bibnamefont {Mukamel}},\ }\bibfield
  {title} {\enquote {\bibinfo {title} {{Exciton-migration and three-pulse
  femtosecond optical spectroscopies of photosynthetic antenna complexes}},}\
  }\href {\doibase 10.1063/1.476212} {\bibfield  {journal} {\bibinfo  {journal}
  {J. Chem. Phys.}\ }\textbf {\bibinfo {volume} {108}},\ \bibinfo {pages}
  {7763} (\bibinfo {year} {1998})}\BibitemShut {NoStop}%
\bibitem [{\citenamefont {Ol{\v{s}}ina}\ \emph {et~al.}(2014)\citenamefont
  {Ol{\v{s}}ina}, \citenamefont {Kramer}, \citenamefont {Kreisbeck},\ and\
  \citenamefont {Man{\v{c}}al}}]{Olsina2014}%
  \BibitemOpen
  \bibfield  {author} {\bibinfo {author} {\bibfnamefont {J.}~\bibnamefont
  {Ol{\v{s}}ina}}, \bibinfo {author} {\bibfnamefont {T.}~\bibnamefont
  {Kramer}}, \bibinfo {author} {\bibfnamefont {C.}~\bibnamefont {Kreisbeck}}, \
  and\ \bibinfo {author} {\bibfnamefont {T.}~\bibnamefont {Man{\v{c}}al}},\
  }\bibfield  {title} {\enquote {\bibinfo {title} {{Exact stochastic unraveling
  of an optical coherence dynamics by cumulant expansion}},}\ }\href {\doibase
  10.1063/1.4898354} {\bibfield  {journal} {\bibinfo  {journal} {The Journal of
  Chemical Physics}\ }\textbf {\bibinfo {volume} {141}},\ \bibinfo {pages}
  {164109} (\bibinfo {year} {2014})},\ \Eprint {http://arxiv.org/abs/1309.0749}
  {arXiv:1309.0749} \BibitemShut {NoStop}%
\bibitem [{\citenamefont {Mascoli}\ \emph {et~al.}(2020)\citenamefont
  {Mascoli}, \citenamefont {Novoderezhkin}, \citenamefont {Liguori},
  \citenamefont {Xu},\ and\ \citenamefont {Croce}}]{Mascoli2020}%
  \BibitemOpen
  \bibfield  {author} {\bibinfo {author} {\bibfnamefont {V.}~\bibnamefont
  {Mascoli}}, \bibinfo {author} {\bibfnamefont {V.}~\bibnamefont
  {Novoderezhkin}}, \bibinfo {author} {\bibfnamefont {N.}~\bibnamefont
  {Liguori}}, \bibinfo {author} {\bibfnamefont {P.}~\bibnamefont {Xu}}, \ and\
  \bibinfo {author} {\bibfnamefont {R.}~\bibnamefont {Croce}},\ }\bibfield
  {title} {\enquote {\bibinfo {title} {{Design principles of solar light
  harvesting in plants: Functional architecture of the monomeric antenna
  CP29}},}\ }\href {\doibase 10.1016/j.bbabio.2020.148156} {\bibfield
  {journal} {\bibinfo  {journal} {Biochim. Biophys. Acta - Bioenerg.}\ }\textbf
  {\bibinfo {volume} {1861}},\ \bibinfo {pages} {148156} (\bibinfo {year}
  {2020})}\BibitemShut {NoStop}%
\bibitem [{\citenamefont {Peterman}\ \emph {et~al.}(1997)\citenamefont
  {Peterman}, \citenamefont {Pullerits}, \citenamefont {van Grondelle},\ and\
  \citenamefont {van Amerongen}}]{Peterman1997}%
  \BibitemOpen
  \bibfield  {author} {\bibinfo {author} {\bibfnamefont {E.~J.~G.}\
  \bibnamefont {Peterman}}, \bibinfo {author} {\bibfnamefont {T.}~\bibnamefont
  {Pullerits}}, \bibinfo {author} {\bibfnamefont {R.}~\bibnamefont {van
  Grondelle}}, \ and\ \bibinfo {author} {\bibfnamefont {H.}~\bibnamefont {van
  Amerongen}},\ }\bibfield  {title} {\enquote {\bibinfo {title}
  {{Electron-phonon coupling and vibronic fine structure of Light-Harvesting
  Complex II of green plants: temperature dependent absorption and
  high-resolution fluorescence spectroscopy}},}\ }\href {\doibase
  10.1021/jp962338e} {\bibfield  {journal} {\bibinfo  {journal} {J. Phys. Chem.
  B}\ }\textbf {\bibinfo {volume} {101}},\ \bibinfo {pages} {4448--4457}
  (\bibinfo {year} {1997})}\BibitemShut {NoStop}%
\bibitem [{\citenamefont {Jurinovich}\ \emph {et~al.}(2015)\citenamefont
  {Jurinovich}, \citenamefont {Viani}, \citenamefont {Prandi}, \citenamefont
  {Renger},\ and\ \citenamefont {Mennucci}}]{Jurinovich2015}%
  \BibitemOpen
  \bibfield  {author} {\bibinfo {author} {\bibfnamefont {S.}~\bibnamefont
  {Jurinovich}}, \bibinfo {author} {\bibfnamefont {L.}~\bibnamefont {Viani}},
  \bibinfo {author} {\bibfnamefont {I.~G.}\ \bibnamefont {Prandi}}, \bibinfo
  {author} {\bibfnamefont {T.}~\bibnamefont {Renger}}, \ and\ \bibinfo {author}
  {\bibfnamefont {B.}~\bibnamefont {Mennucci}},\ }\bibfield  {title} {\enquote
  {\bibinfo {title} {{Towards an ab initio description of optical spectra of
  light-harvesting antennae: Application to the CP29 complex of Photosystem
  II}},}\ }\href {\doibase 10.1039/C4CP05647G} {\bibfield  {journal} {\bibinfo
  {journal} {Phys. Chem. Chem. Phys.}\ }\textbf {\bibinfo {volume} {17}},\
  \bibinfo {pages} {14405--14416} (\bibinfo {year} {2015})}\BibitemShut
  {NoStop}%
\bibitem [{\citenamefont {M{\"{u}}h}\ \emph {et~al.}(2014)\citenamefont
  {M{\"{u}}h}, \citenamefont {Lindorfer}, \citenamefont {{Schmidt am Busch}},\
  and\ \citenamefont {Renger}}]{Muh2014}%
  \BibitemOpen
  \bibfield  {author} {\bibinfo {author} {\bibfnamefont {F.}~\bibnamefont
  {M{\"{u}}h}}, \bibinfo {author} {\bibfnamefont {D.}~\bibnamefont
  {Lindorfer}}, \bibinfo {author} {\bibfnamefont {M.}~\bibnamefont {{Schmidt am
  Busch}}}, \ and\ \bibinfo {author} {\bibfnamefont {T.}~\bibnamefont
  {Renger}},\ }\bibfield  {title} {\enquote {\bibinfo {title} {{Towards a
  structure-based exciton Hamiltonian for the CP29 antenna of photosystem
  II}},}\ }\href {\doibase 10.1039/C3CP55166K} {\bibfield  {journal} {\bibinfo
  {journal} {Phys. Chem. Chem. Phys.}\ }\textbf {\bibinfo {volume} {16}},\
  \bibinfo {pages} {11848--11863} (\bibinfo {year} {2014})}\BibitemShut
  {NoStop}%
\bibitem [{\citenamefont {Jennings}\ \emph {et~al.}(2003)\citenamefont
  {Jennings}, \citenamefont {Zucchelli}, \citenamefont {Croce},\ and\
  \citenamefont {Garlaschi}}]{Jennings2003}%
  \BibitemOpen
  \bibfield  {author} {\bibinfo {author} {\bibfnamefont {R.~C.}\ \bibnamefont
  {Jennings}}, \bibinfo {author} {\bibfnamefont {G.}~\bibnamefont {Zucchelli}},
  \bibinfo {author} {\bibfnamefont {R.}~\bibnamefont {Croce}}, \ and\ \bibinfo
  {author} {\bibfnamefont {F.~M.}\ \bibnamefont {Garlaschi}},\ }\bibfield
  {title} {\enquote {\bibinfo {title} {{The photochemical trapping rate from
  red spectral states in PSI-LHCI is determined by thermal activation of energy
  transfer to bulk chlorophylls}},}\ }\href {\doibase
  10.1016/S0005-2728(02)00399-7} {\bibfield  {journal} {\bibinfo  {journal}
  {Biochim. Biophys. Acta - Bioenerg.}\ }\textbf {\bibinfo {volume} {1557}},\
  \bibinfo {pages} {91--98} (\bibinfo {year} {2003})}\BibitemShut {NoStop}%
\bibitem [{\citenamefont {Kloeden}\ and\ \citenamefont
  {Platen}(2011)}]{kloeden2011numerical}%
  \BibitemOpen
  \bibfield  {author} {\bibinfo {author} {\bibfnamefont {P.~E.}\ \bibnamefont
  {Kloeden}}\ and\ \bibinfo {author} {\bibfnamefont {E.}~\bibnamefont
  {Platen}},\ }\href {https://books.google.co.za/books?id=BCvtssom1CMC} {\emph
  {\bibinfo {title} {{Numerical Solution of Stochastic Differential
  Equations}}}},\ Stochastic Modelling and Applied Probability\ (\bibinfo
  {publisher} {Springer Berlin Heidelberg},\ \bibinfo {year}
  {2011})\BibitemShut {NoStop}%
\bibitem [{\citenamefont {Fassioli}\ \emph {et~al.}(2014)\citenamefont
  {Fassioli}, \citenamefont {Dinshaw}, \citenamefont {Arpin},\ and\
  \citenamefont {Scholes}}]{Fassioli2014a}%
  \BibitemOpen
  \bibfield  {author} {\bibinfo {author} {\bibfnamefont {F.}~\bibnamefont
  {Fassioli}}, \bibinfo {author} {\bibfnamefont {R.}~\bibnamefont {Dinshaw}},
  \bibinfo {author} {\bibfnamefont {P.~C.}\ \bibnamefont {Arpin}}, \ and\
  \bibinfo {author} {\bibfnamefont {G.~D.}\ \bibnamefont {Scholes}},\
  }\bibfield  {title} {\enquote {\bibinfo {title} {{Photosynthetic light
  harvesting: excitons and coherence}},}\ }\href {\doibase
  10.1098/rsif.2013.0901} {\bibfield  {journal} {\bibinfo  {journal} {J. R.
  Soc. Interface}\ }\textbf {\bibinfo {volume} {11}},\ \bibinfo {pages}
  {20130901} (\bibinfo {year} {2014})}\BibitemShut {NoStop}%
\bibitem [{\citenamefont {Lindorfer}, \citenamefont {M{\"{u}}h},\ and\
  \citenamefont {Renger}(2017)}]{Lindorfer2017}%
  \BibitemOpen
  \bibfield  {author} {\bibinfo {author} {\bibfnamefont {D.}~\bibnamefont
  {Lindorfer}}, \bibinfo {author} {\bibfnamefont {F.}~\bibnamefont
  {M{\"{u}}h}}, \ and\ \bibinfo {author} {\bibfnamefont {T.}~\bibnamefont
  {Renger}},\ }\bibfield  {title} {\enquote {\bibinfo {title} {{Origin of
  non-conservative circular dichroism of the CP29 antenna complex of
  Photosystem II}},}\ }\href {\doibase 10.1039/C6CP08778G} {\bibfield
  {journal} {\bibinfo  {journal} {Phys. Chem. Chem. Phys.}\ }\textbf {\bibinfo
  {volume} {19}},\ \bibinfo {pages} {7524--7536} (\bibinfo {year}
  {2017})}\BibitemShut {NoStop}%
\bibitem [{\citenamefont {Novoderezhkin}\ \emph {et~al.}(2010)\citenamefont
  {Novoderezhkin}, \citenamefont {Doust}, \citenamefont {Curutchet},
  \citenamefont {Scholes},\ and\ \citenamefont {van Grondelle}}]{Novod_PCE545}%
  \BibitemOpen
  \bibfield  {author} {\bibinfo {author} {\bibfnamefont {V.~I.}\ \bibnamefont
  {Novoderezhkin}}, \bibinfo {author} {\bibfnamefont {A.~B.}\ \bibnamefont
  {Doust}}, \bibinfo {author} {\bibfnamefont {C.}~\bibnamefont {Curutchet}},
  \bibinfo {author} {\bibfnamefont {G.~D.}\ \bibnamefont {Scholes}}, \ and\
  \bibinfo {author} {\bibfnamefont {R.}~\bibnamefont {van Grondelle}},\
  }\bibfield  {title} {\enquote {\bibinfo {title} {{Excitation dynamics in
  Phycoerythrin 545: Modeling of steady-state spectra and transient absorption
  with modified Redfield theory}},}\ }\href {\doibase
  10.1016/j.bpj.2010.04.039} {\bibfield  {journal} {\bibinfo  {journal}
  {Biophys. J.}\ }\textbf {\bibinfo {volume} {99}},\ \bibinfo {pages}
  {344--352} (\bibinfo {year} {2010})}\BibitemShut {NoStop}%
\bibitem [{\citenamefont {van Grondelle}\ and\ \citenamefont
  {Novoderezhkin}(2006)}]{VanGrondelle2006}%
  \BibitemOpen
  \bibfield  {author} {\bibinfo {author} {\bibfnamefont {R.}~\bibnamefont {van
  Grondelle}}\ and\ \bibinfo {author} {\bibfnamefont {V.~I.}\ \bibnamefont
  {Novoderezhkin}},\ }\bibfield  {title} {\enquote {\bibinfo {title} {{Energy
  transfer in photosynthesis: experimental insights and quantitative
  models}},}\ }\href {\doibase 10.1039/B514032C} {\bibfield  {journal}
  {\bibinfo  {journal} {Phys. Chem. Chem. Phys.}\ }\textbf {\bibinfo {volume}
  {8}},\ \bibinfo {pages} {793--807} (\bibinfo {year} {2006})}\BibitemShut
  {NoStop}%
\bibitem [{\citenamefont {Jun}\ \emph {et~al.}(2021)\citenamefont {Jun},
  \citenamefont {Yang}, \citenamefont {Choi}, \citenamefont {Isaji},
  \citenamefont {Tamiaki}, \citenamefont {Ihee},\ and\ \citenamefont
  {Kim}}]{D1CP03413H}%
  \BibitemOpen
  \bibfield  {author} {\bibinfo {author} {\bibfnamefont {S.}~\bibnamefont
  {Jun}}, \bibinfo {author} {\bibfnamefont {C.}~\bibnamefont {Yang}}, \bibinfo
  {author} {\bibfnamefont {S.}~\bibnamefont {Choi}}, \bibinfo {author}
  {\bibfnamefont {M.}~\bibnamefont {Isaji}}, \bibinfo {author} {\bibfnamefont
  {H.}~\bibnamefont {Tamiaki}}, \bibinfo {author} {\bibfnamefont
  {H.}~\bibnamefont {Ihee}}, \ and\ \bibinfo {author} {\bibfnamefont
  {J.}~\bibnamefont {Kim}},\ }\bibfield  {title} {\enquote {\bibinfo {title}
  {{Exciton delocalization length in chlorosomes investigated by lineshape
  dynamics of two-dimensional electronic spectra}},}\ }\href {\doibase
  10.1039/D1CP03413H} {\bibfield  {journal} {\bibinfo  {journal} {Phys. Chem.
  Chem. Phys.}\ }\textbf {\bibinfo {volume} {23}},\ \bibinfo {pages}
  {24111--24117} (\bibinfo {year} {2021})}\BibitemShut {NoStop}%
\bibitem [{\citenamefont {Zucchelli}\ \emph {et~al.}(1994)\citenamefont
  {Zucchelli}, \citenamefont {Jennings}, \citenamefont {Garlaschi},
  \citenamefont {Dainese}, \citenamefont {Breton},\ and\ \citenamefont
  {Bassi}}]{Zucchelli1994}%
  \BibitemOpen
  \bibfield  {author} {\bibinfo {author} {\bibfnamefont {G.}~\bibnamefont
  {Zucchelli}}, \bibinfo {author} {\bibfnamefont {R.~C.}\ \bibnamefont
  {Jennings}}, \bibinfo {author} {\bibfnamefont {F.~M.}\ \bibnamefont
  {Garlaschi}}, \bibinfo {author} {\bibfnamefont {P.}~\bibnamefont {Dainese}},
  \bibinfo {author} {\bibfnamefont {J.}~\bibnamefont {Breton}}, \ and\ \bibinfo
  {author} {\bibfnamefont {R.}~\bibnamefont {Bassi}},\ }\bibfield  {title}
  {\enquote {\bibinfo {title} {{Gaussian decomposition of absorption and linear
  dichroism spectra of outer antenna complexes of Photosystem II}},}\ }\href
  {\doibase 10.1021/bi00196a016} {\bibfield  {journal} {\bibinfo  {journal}
  {Biochemistry}\ }\textbf {\bibinfo {volume} {33}},\ \bibinfo {pages}
  {8982--8990} (\bibinfo {year} {1994})}\BibitemShut {NoStop}%
\end{thebibliography}%

\end{document}



\title{Supplementary Information\\ \vspace{0.6 cm}
Accuracy of approximate methods for the calculation of absorption-type linear spectra with a complex system-bath coupling} 



\author{J.A. N\"{o}thling}
\affiliation{Department of Physics, University of Pretoria, 0002 Pretoria, South Africa}
\affiliation{National Institute for Theoretical and Computational Sciences (NITheCS), South Africa}
 \author{Tom\'{a}\v{s} Man\v{c}al}
\affiliation{Faculty of Mathematics and Physics, Charles University, Ke Karlovu 5, CZ-121 16 Prague 2, Czech Republic}
 \author{T.P.J. Kr\"{uger}}
\affiliation{Department of Physics, University of Pretoria, 0002 Pretoria, South Africa}
\affiliation{National Institute for Theoretical and Computational Sciences (NITheCS), South Africa}

\maketitle 

\makeatletter 
\renewcommand{\thefigure}{S\@arabic\c@figure}
\makeatother

\begin{figure*}[!htbp]
\includegraphics[scale=1.1, trim=0 1cm 0 1cm, clip]{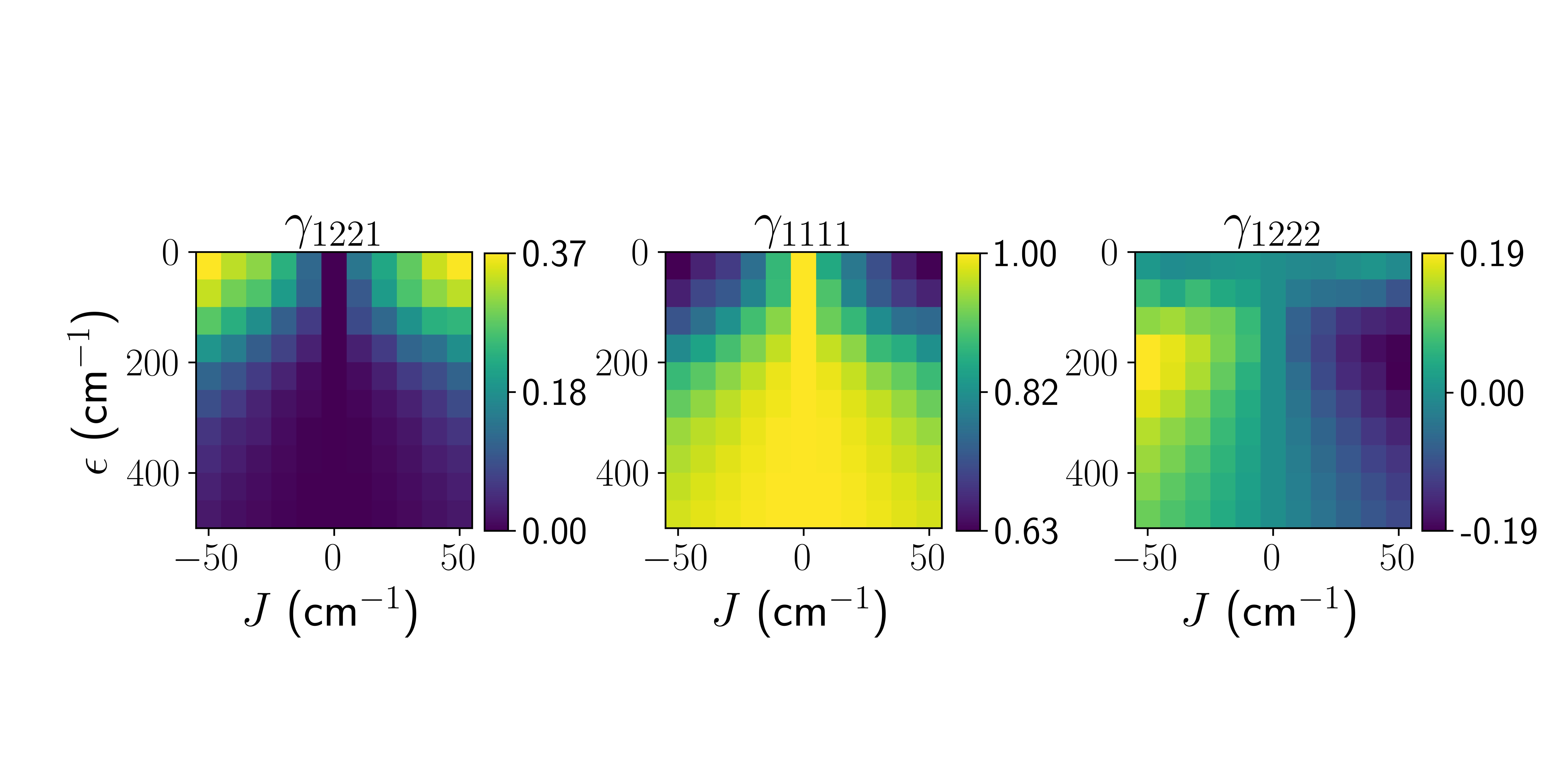}
\caption{The participation factors $\gamma_{1221} = \sum_{n=1}^2(c^{n1})^2(c^{n2})^2$ (left), $\gamma_{1111} = \sum_{n=1}^2(c^{n1})^4$ (middle), and $\gamma_{1222} = \sum_{n=1}^2c^{n1}(c^{n2})^3$ (right), where $(c^{n\alpha})^2$ is the participation ratio of pigment $n$ to the exciton state $\alpha$. The participation factors $\gamma$ were averaged over 200 realizations of disorder with $\sigma_{\text{FWHM}}=140\pcm$.}
\end{figure*}

\begin{figure*}[!htbp]
\includegraphics[scale=1]{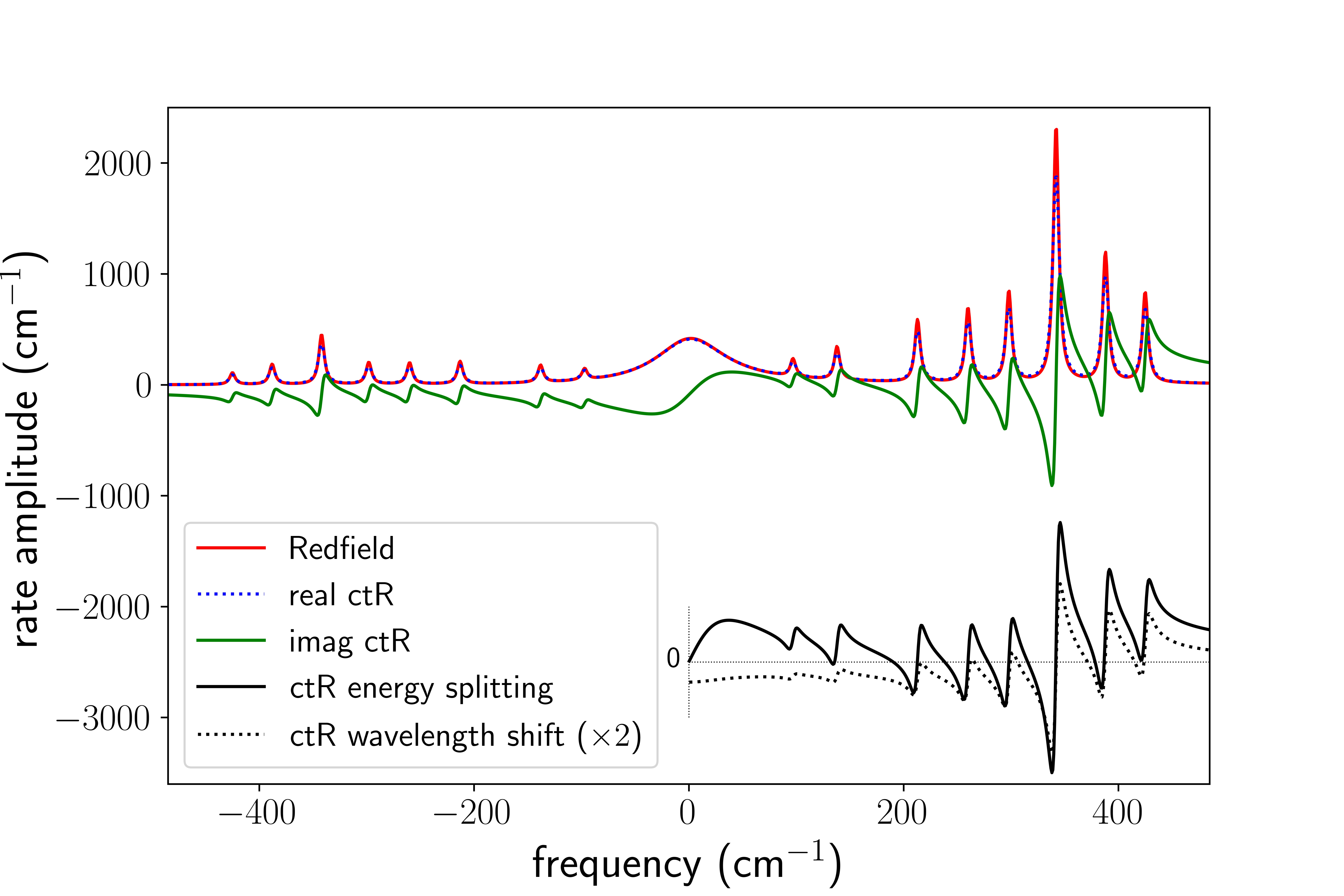}
\caption{Redfield rates and the real and imaginary parts of the long-time ctR rates as functions of the excitonic energy gap $\omega_{\alpha\beta}$. The long-time ctR rates are calculated as $\xi_l(\omega) = \xi(t_l, \omega)/t_l$, where $\xi(t)$ is the coherent time-dependent Redfield term and $t_l$ is a large finite time. Note that the rates were not corrected for pigment participation to excitons. The total (participation-uncorrected) splitting in a dimer spectrum is given by $\text{Im}\,\xi_l(\omega) - \text{Im}\,\xi_l(-\omega)$ and the average shift is given by  $\frac{1}{2}\Bigl(\text{Im}\,\xi_l(\omega) + \text{Im}\,\xi_l(-\omega)\Bigr)$.}
\end{figure*}

\begin{figure*}[!htbp]
\includegraphics[scale=1]{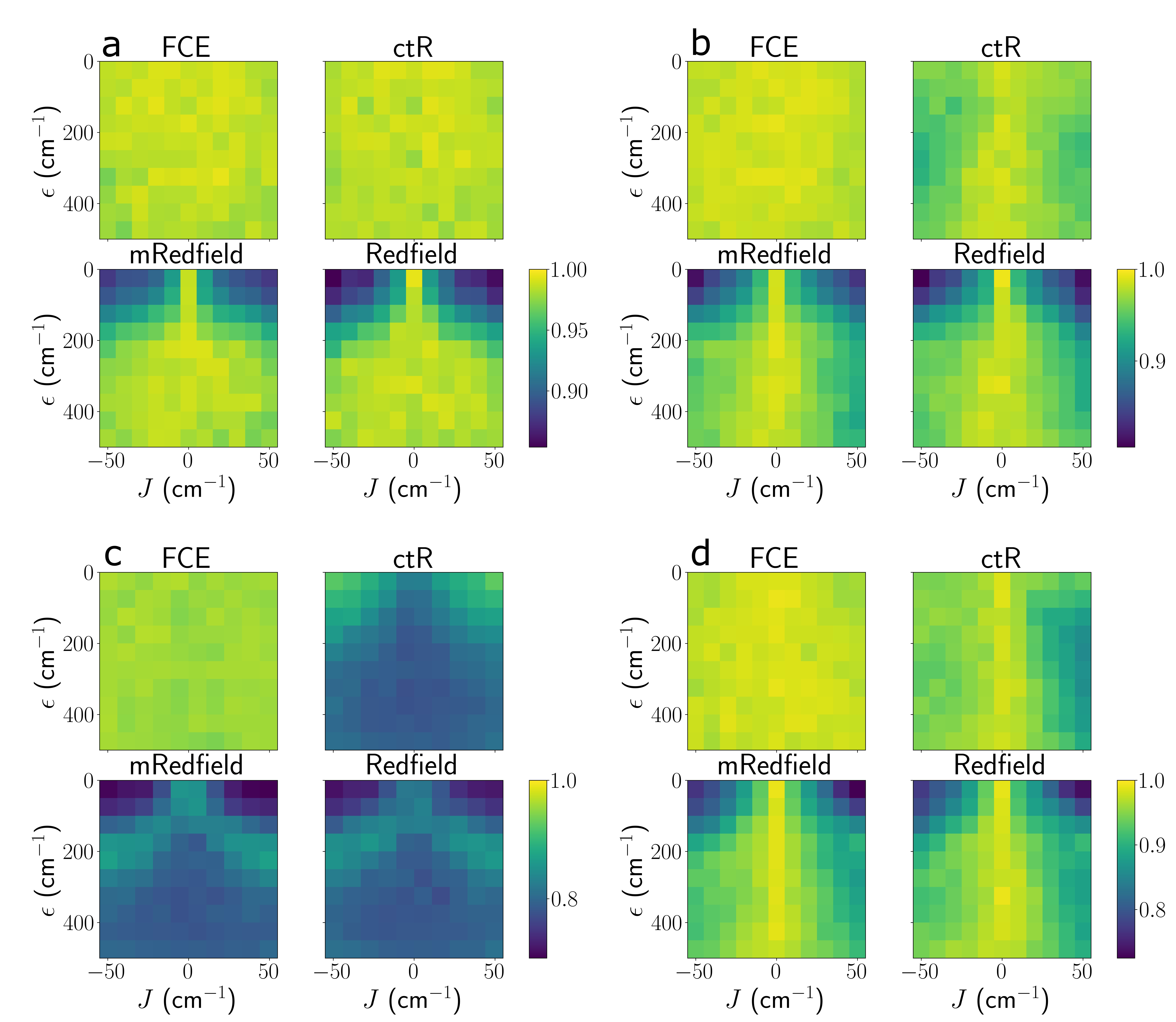}
\caption{Dependence of the quality of absorption-type spectra on the site energy and coupling at $100 \text{ K}$. Qualities are shown for the absorption dipole factors (\textbf{a}) $f^{1,0}_{0,1}$ and (\textbf{b}) $f^{1,1}_{1,1}$, (\textbf{c}) the CD dipole factor  $f^{0,1}_{1,0}$, and (\textbf{d}) the LD dipole factor  $f^{1,-1}_{-1,0}$. A disorder of $\sigma_{\text{FWHM}}=140\pcm$ was used for the calculation of approximate spectra.}
\end{figure*}

\begin{figure*}[!htbp]
\includegraphics[scale=1.2]{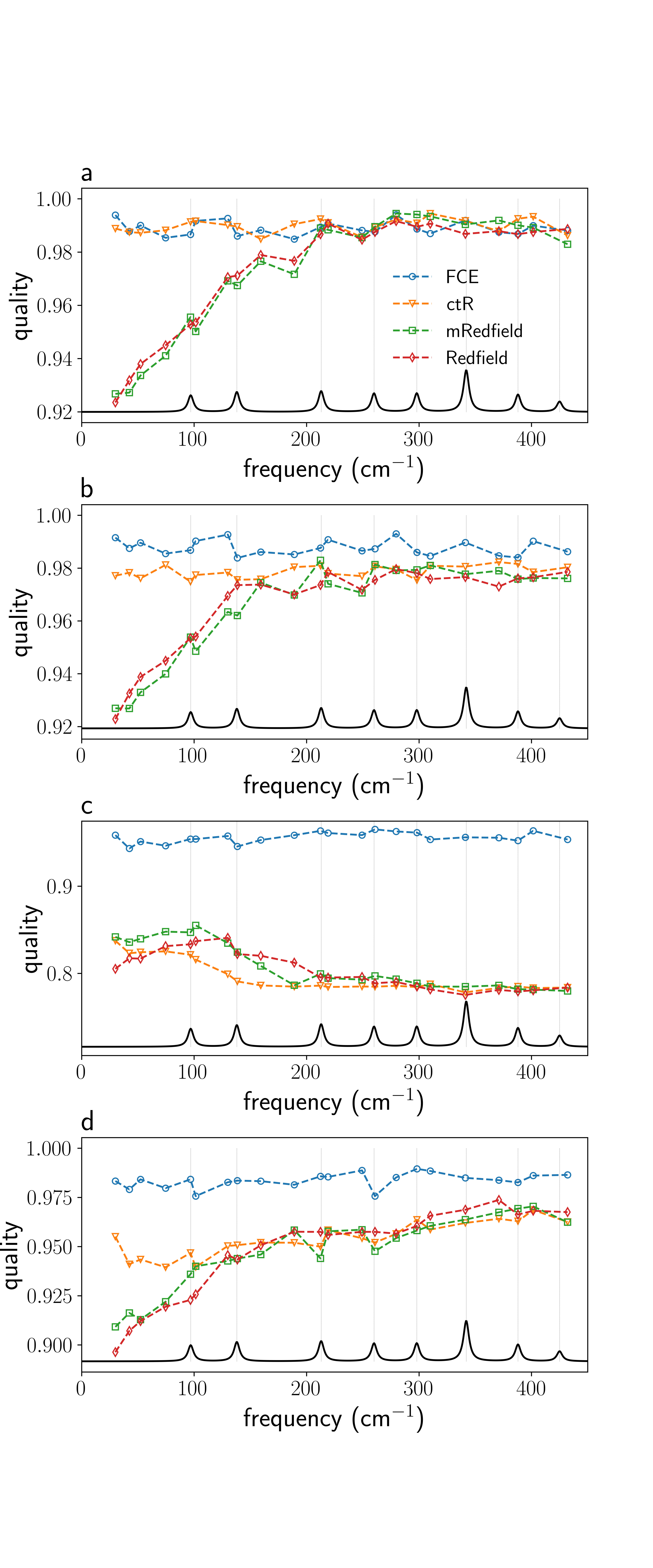}
\caption{Quality of absorption-type spectra at 100 K as a function of the excitonic energy gap for $J=15\pcm$.  Qualities are shown for the absorption dipole factors (\textbf{a}) $f^{1,0}_{0,1}$ and (\textbf{b}) $f^{1,1}_{1,1}$, (\textbf{c}) CD dipole factor  $f^{0,1}_{1,0}$, and (\textbf{d}) LD dipole factor $f^{1,-1}_{-1,0}$. A disorder of $\sigma_{\text{FWHM}}=140\pcm$ was used for the calculation of approximate spectra.}
\end{figure*}

\begin{figure*}[!htbp]
\includegraphics[scale=1.2]{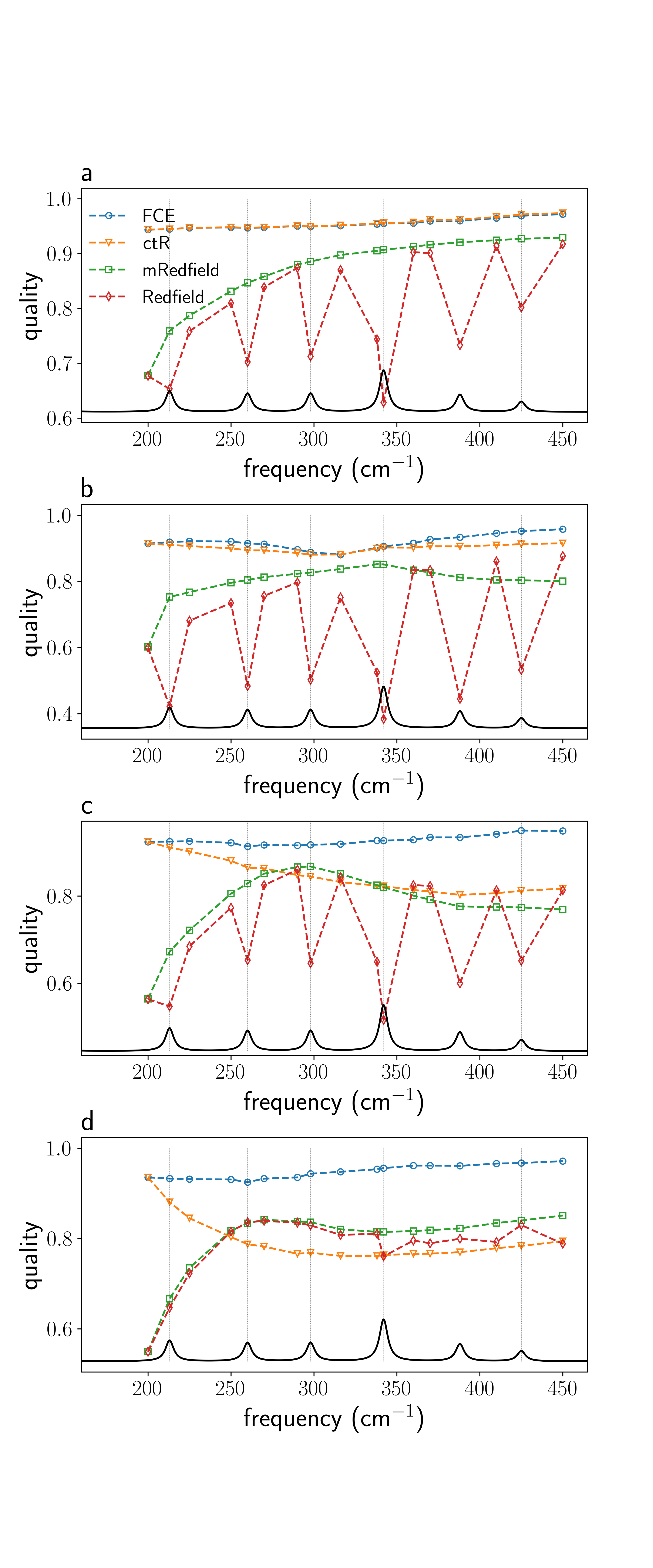}
\caption{Quality of absorption-type spectra for an excitonic coupling of $J=100\pcm$ at 100 K as a function of the excitonic energy gap.  Qualities are shown for the absorption dipole factors (\textbf{a}) $f^{1,0}_{0,1}$ and (\textbf{b}) $f^{1,1}_{1,1}$, (\textbf{c}) CD dipole factor  $f^{0,1}_{1,0}$, and (\textbf{d}) LD dipole factor $f^{1,-1}_{-1,0}$. No disorder was included for the calculation of approximate spectra.}
\end{figure*}

\begin{figure*}[!htbp]
\includegraphics[scale=1.2]{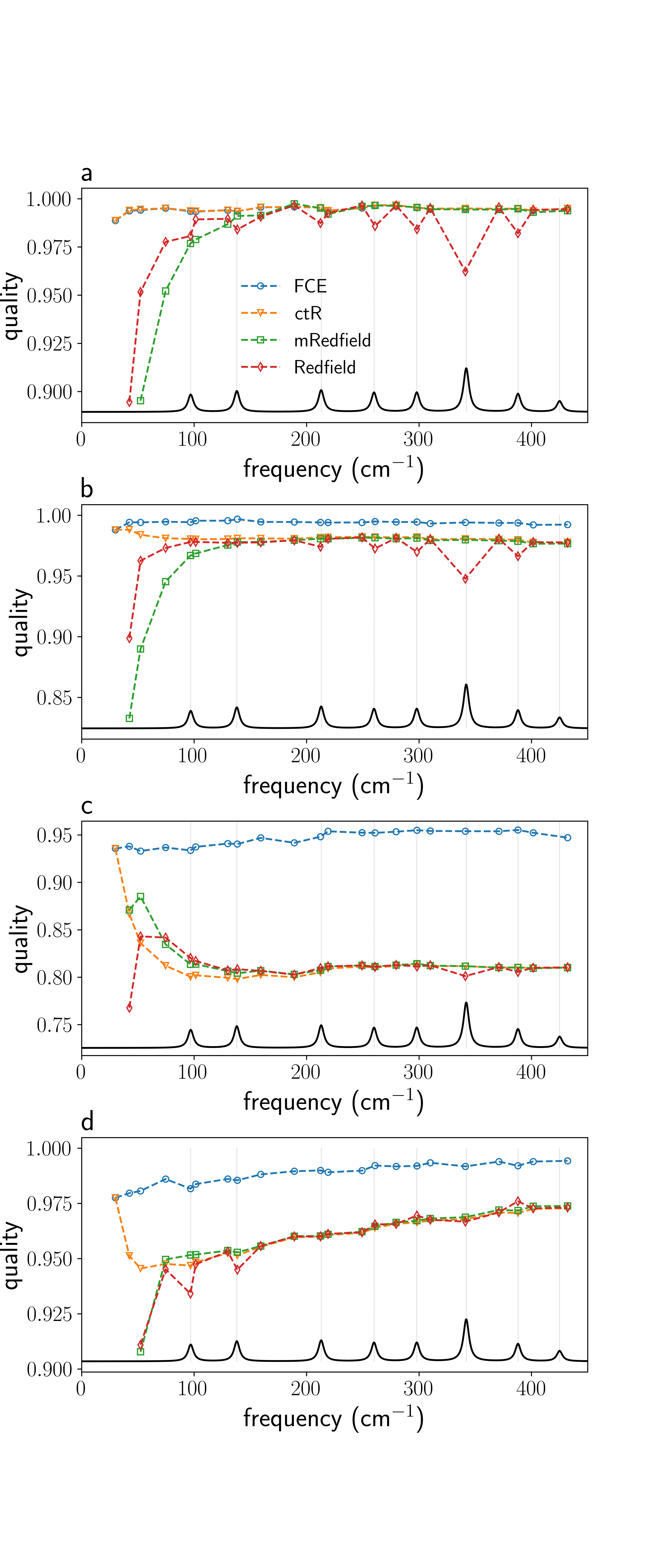}
\caption{Quality of absorption-type spectra for an excitonic coupling of $J=15\pcm$ at 300 K as a function of the excitonic energy gap.  Qualities are shown for the absorption dipole factors (\textbf{a}) $f^{1,0}_{0,1}$ and (\textbf{b}) $f^{1,1}_{1,1}$, (\textbf{c}) CD dipole factor $f^{0,1}_{1,0}$, and (\textbf{d}) LD dipole factor $f^{1,-1}_{-1,0}$. No disorder was included for the calculation of approximate spectra.}
\end{figure*}

\begin{figure*}[!htbp]
\includegraphics[scale=1, trim=0 1cm 0 1cm, clip]{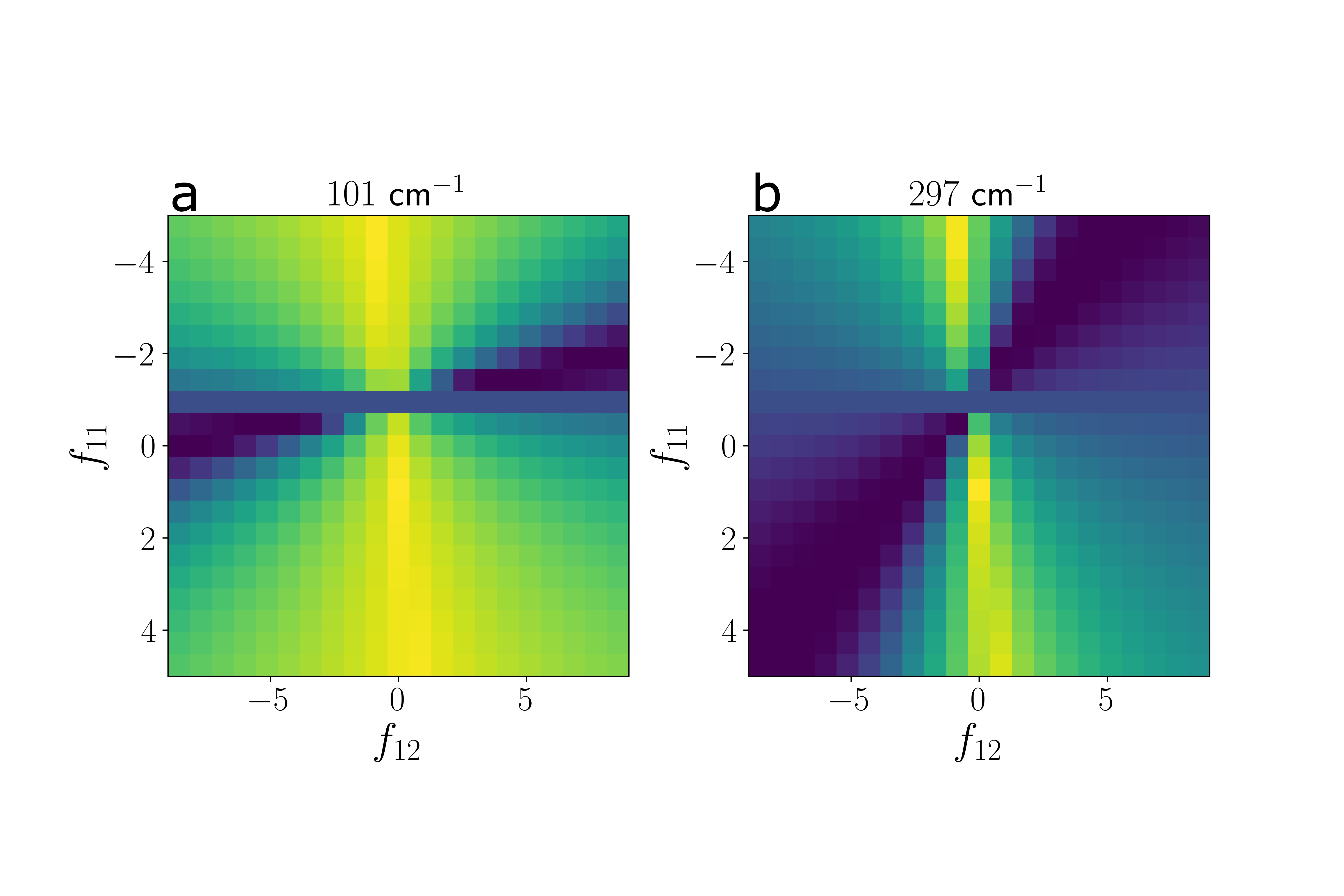}
\caption{Simulation of the accuracy of the secular approximation for dipole factors of the form $f^{f_{11},f_{12}}_{f_{12}, \, 1}$, showing the relative contribution of the diagonal elements of a $2\times2$ matrix $I_{\text{rep}}$ that represents the full spectral tensor, to the sum prescribed by the dipole factor $f$. We chose the elementwise absolute maxima (taken along the time dimension) (\textbf{a}) $\begin{bmatrix} 1 & 0.055 \\ 0.055 & 1 \end{bmatrix}$ and (\textbf{b}) $\begin{bmatrix} 1 & 0.33 \\ 0.33 & 1  \end{bmatrix}$ as representative matrices for the absorption tensors corresponding to the site-basis Hamiltonians (\textbf{a}) $\begin{bmatrix} 97 & 15 \\ 15 & 0 \end{bmatrix}$ and (\textbf{b}) $\begin{bmatrix} 298 & 100 \\ 100 & 0 \end{bmatrix}$. The exciton gaps for these Hamiltonians are indicated on top. Differences between these simulations and the true accuracy of the secular approximation for real spectra may arise due to the impossibility of summarizing a time-dependent tensor fully in a $2\times2$ matrix. A nonlinear transformation was applied (the same transformation for all dipole factors and both Hamiltonians) to represent the diagonal contribution graphically.} 
\end{figure*}

\begin{figure*}[!htbp]
\includegraphics[scale=0.7]{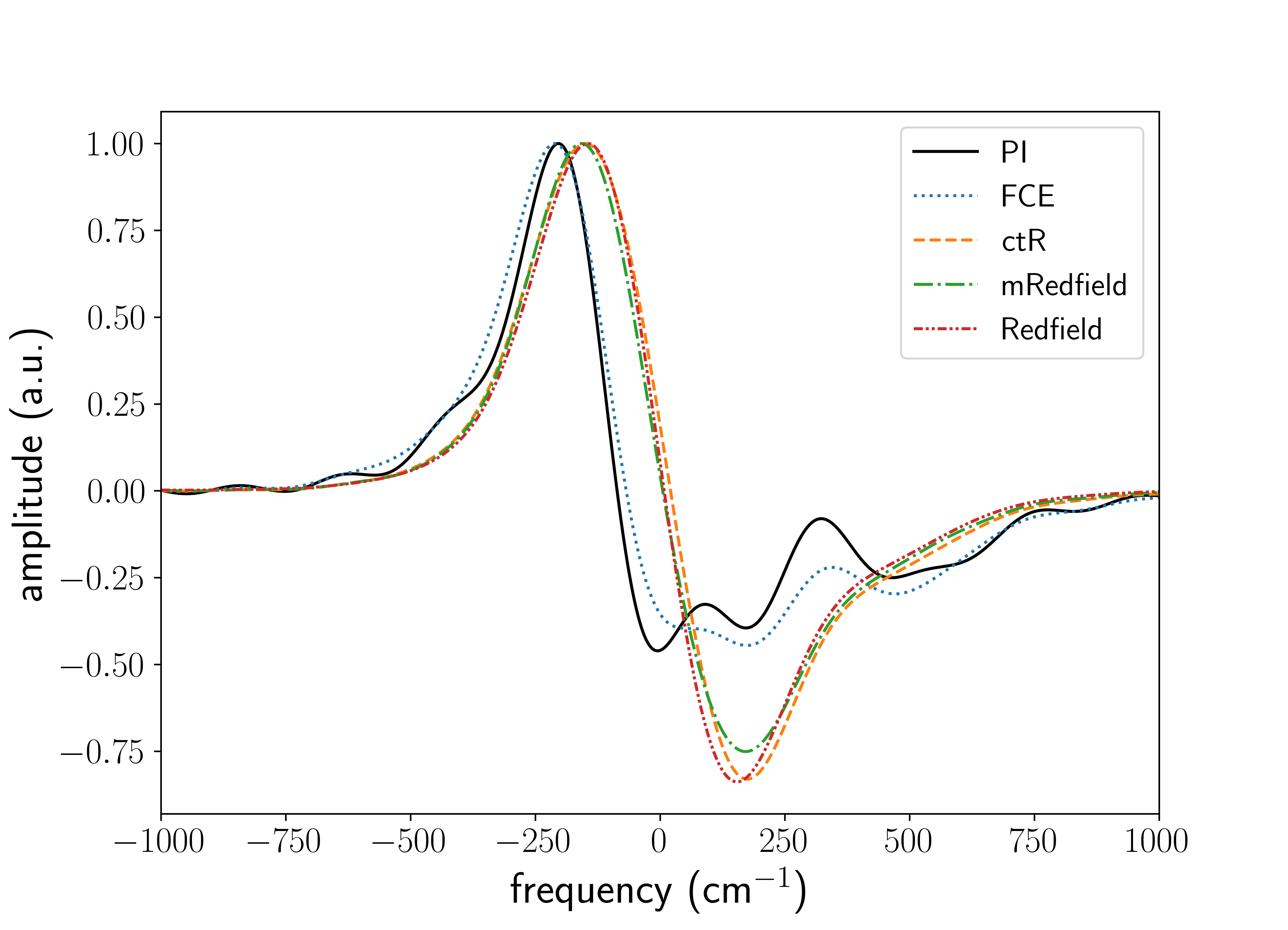}
\caption{The spectra for the dipole factor $f^{-1,3}_{3,1}$. Note the contribution of vibronic coherence to the PI and FCE spectra, but not to the secular spectra. The small oscillations in the PI spectrum at large frequencies are due to incomplete decay of the PI absorption tensor.}
\end{figure*}

\begin{figure*}[!htbp]
\includegraphics[scale=1.2]{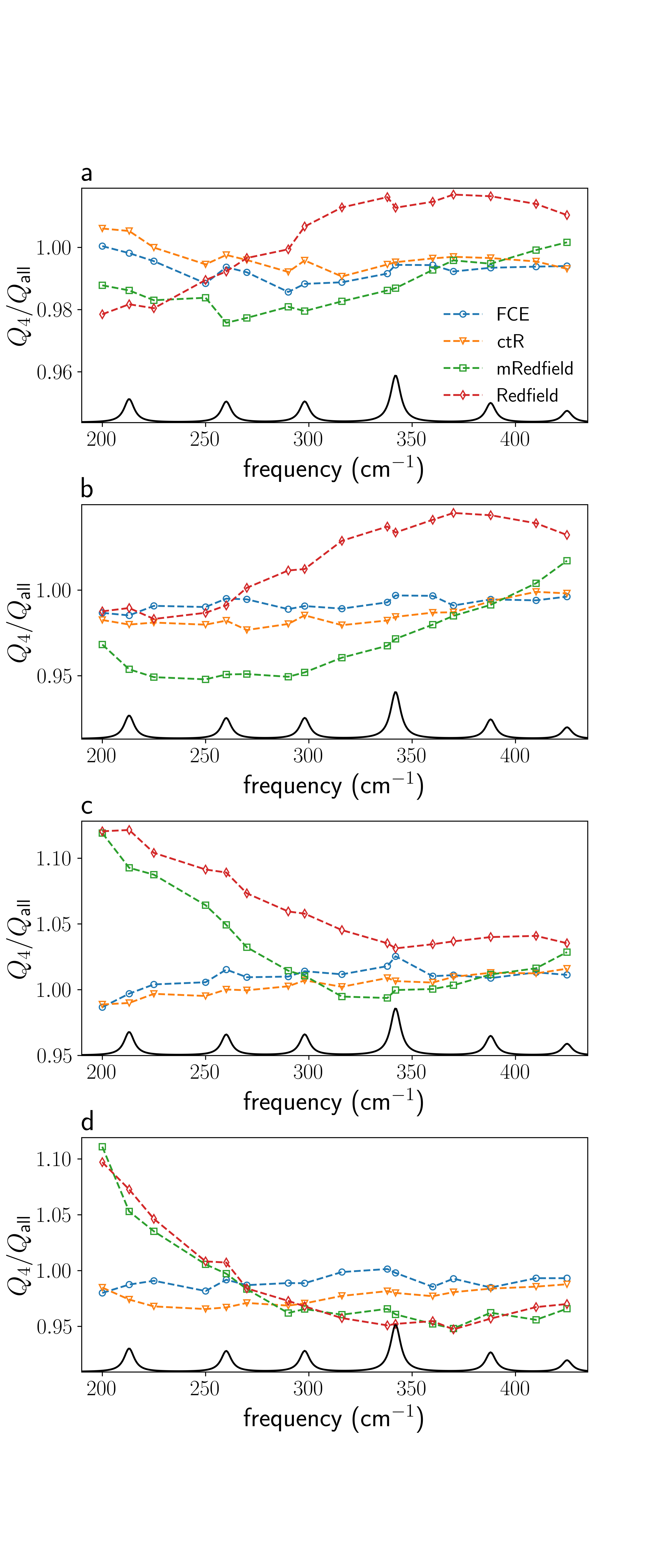}
\caption{The ratio of qualities obtained when only the fourth intramolecular mode was included in the spectral density and when all eight modes were included. Qualities are shown for the absorption dipole factors (\textbf{a}) $f^{1,0}_{0,1}$ and (\textbf{b}) $f^{1,1}_{1,1}$, (\textbf{c}) CD dipole factor  $f^{0,1}_{1,0}$, and (\textbf{d}) LD dipole factor $f^{1,-1}_{-1,0}$. A disorder of $\sigma_{\text{FWHM}}=140\pcm$ was used for the calculation of approximate spectra.}
\end{figure*}

\clearpage
\bibliography{library}